\documentclass[prd,aps,twocolumn,amsmath,amssymb,nofootinbib,preprintnumbers]
{revtex4}

\usepackage{graphicx}
\usepackage{dcolumn}
\usepackage{bm}
\usepackage{amsmath}
\usepackage{amsfonts}
\usepackage{euscript,bbm}
\usepackage{ifthen}
\usepackage{psfrag}

\def\ls{\mathrel{\lower4pt\vbox{\lineskip=0pt\baselineskip=0pt
           \hbox{$<$}\hbox{$\sim$}}}}
\def\gs{\mathrel{\lower4pt\vbox{\lineskip=0pt\baselineskip=0pt
           \hbox{$>$}\hbox{$\sim$}}}}
\def\drawbox#1#2{\hrule height#2pt

\hbox{\vrule width#2pt height#1pt \kern#1pt
              \vrule width#2pt}
              \hrule height#2pt}

\def\Asym#1#2{\vcenter{\vbox{\drawbox{#1}{#2}
              \kern-#2pt       
              \drawbox{#1}{#2}}}}

\newcommand{\be}{\begin{equation}}
\newcommand{\ee}{\end{equation}}
\newcommand{\bea}{\begin{eqnarray}}
\newcommand{\eea}{\end{eqnarray}}
\newcommand{\sel}{\ensuremath{\tilde{e}}}

\newcommand{\neu}[1]{\ensuremath{\tilde{\chi}_{#1}^0}}

\newcommand{\chpm}[1]{\ensuremath{\tilde{\chi}_{#1}^{\pm}}}
\newcommand{\sq}{\ensuremath{\tilde{q}}}
\newcommand{\stau}{\ensuremath{\tilde{\tau}}}
\newcommand{\sg}{\ensuremath{\tilde{g}}}
\newcommand{\sd}{\ensuremath{\tilde{d}}}
\newcommand{\su}{\ensuremath{\tilde{u}}}
\newcommand{\st}{\ensuremath{\tilde{t}}}

\newcommand{\met} {{E\!\!\!\!/_{\rm T}}}

\newcommand{ \pythia } {{\tt PYTHIA}}
\newcommand{ \isasugra } {{\tt ISASUGRA}}
\newcommand{ \isajet }    {{\tt ISAJET}}
\newcommand{ \pgs }    {{\tt PGS4}}
\newcommand{ \darksusy }    {{\tt darkSUSY}}

\begin{document}

%
\title{Diagnosis of Supersymmetry Breaking Mediation Schemes by Mass Reconstruction at the LHC}

\author{Bhaskar Dutta$^{1}$}
\author{Teruki Kamon$^{1,2}$}
\author{Abram Krislock$^{1,3}$}
\author{Kuver Sinha$^{1}$}
\author{Kechen Wang$^{1}$}

\affiliation{$^{1}$~Department of Physics \& Astronomy, Mitchell Institute for Fundamental Physics, Texas A\&M University, College Station, TX 77843-4242, USA \\
$^{2}$~Department of Physics, Kyungpook National University, Daegu 702-701, South Korea \\
$^{3}$~Department of Physics, AlbaNova, Stockholm University, SE-106 91, Stockholm, Sweden}

\begin{abstract}

If supersymmetry is discovered at the LHC, the next question will be the determination of the underlying model. While this may be challenging or even intractable, a more optimistic question is whether we can understand the main contours of any particular paradigm of the mediation of supersymmetry breaking. The determination of superpartner masses through endpoint measurements of kinematic observables arising from cascade decays is a powerful diagnostic tool. In particular, the determination of the gaugino sector has the potential to discriminate between certain mediation schemes (not all schemes, and not between different UV realizations of a given scheme). We reconstruct gaugino masses, choosing a model where anomaly contributions to supersymmetry breaking are important (KKLT compactification), and find the gaugino unification scale. Moreover, reconstruction of other superpartner masses allows us to solve for the parameters defining the UV model. The analysis is performed in the stop and stau coannihilation regions where the lightest neutralinos are mainly gauginos, to additionally satisfy dark matter constraints. We thus develop observables to determine stau and stop masses to verify that the coannihilation mechanism is indeed operational, and solve for the relic density.


\end{abstract}
MIFPA-11-56

\maketitle



\section{Introduction}\label{intro}

The Large Hadron Collider (LHC) will soon be testing many ideas for physics beyond the Standard Model (SM). Of these, low energy supersymmetry \cite{Martin:1997ns} is the best motivated candidate for new TeV scale physics.

If supersymmetric partners are indeed detected at the LHC, the next step would be understanding the underlying scheme by which supersymmetry breaking is mediated to the visible sector. Of course, the full determination of an exact model of supersymmetry breaking and mediation may be challenging, but one can hope that the broad contours of the scheme can be understood.

How soon (if at all) will we be able to distinguish the underlying mechanism by which supersymmetry breaking is mediated to the visible sector?

It is useful to first understand the above question from the perspective of distinguishability studies undertaken from a top-down point of view \cite{arXiv:1110.1287}. Broadly, the program is to fit a set of UV parameters defining a sufficiently tractable model of supersymmetry breaking and mediation with observables measured in the experiment, and ask questions about distinguishability in the space of UV model parameters. Some models one can consider in such distinguishability tests are $(i)$ mSUGRA or CMSSM \cite{msugra} (defined at the GUT scale by universal scalar and gaugino masses and trilinear couplings $m_0, m_{1/2}$, and $A_0$, as well as ${\rm tan} \beta$ and ratio of Higgs vevs), $(ii)$ the minimal anomaly mediated supersymmetry model \cite{Giudice:1998bp} (defined by $m_0$, ${\rm tan}\beta$, and universal superpartner mass scale $m_{aux}$), $(iii)$ minimal gauge mediation (defined by the messenger scale $M_{mess}$, supersymmetry breaking scale $\Lambda$ and ${\rm tan}\beta$), as well as some examples inspired by string phenomenology: $(iv)$ Large Volume Scenarios \cite{hep-th/0505076} (defined by $m_0$ and ${\rm tan}\beta$), and $(v)$ mirage mediation models \cite{Choi:2005uz} in the context of KKLT \cite{Kachru:2003aw}.
 
However, the important point is that \textit{a given set of measured observables will in general not only point back to different UV models, it may point to different mediation schemes. } It may be more optimistic to ask whether we can at least distinguish between more gross properties, such as mediation schemes.

It is important, then, to first discern whether a particular scheme of supersymmetry breaking mediation (such as, say, anomaly contributions) shows up unmistakably in the low energy spectrum, before locking into a particular UV model and undertaking a distinguishability study in the UV model parameter space or attempting to constrain it. 

Which masses should one attempt to solve to this end?

One can look for relatively model-independent soft mass patterns that hold out clues for mediation schemes. As emphasised in \cite{Choi:2007ka}, the gaugino sector offers a particularly promising portal for understanding the underlying mechanism of mediation. This is mainly because within the context of the Minimal Supersymmetric extension of the Standard Model (MSSM), the quantity $M_a/g_a^2$ does not run at one loop (where $a=3,2,1$ refers to the SM gauge groups $SU(3) \times SU(2) \times U(1)$, while $M_a$ refer to gaugino masses and $g_a$ to gauge couplings). If one makes two further assumptions: $(i)$ that gaugino masses are dominated by contributions determined by tree level gauge kinetic functions and $(ii)$ these tree level contributions are universal, then one obtains the well known mSUGRA pattern, with gaugino unification at the GUT scale 
\be \label{mSUGRApattern}
{\rm mSUGRA:} \,\,\,\,\, M_1:M_2:M_3 \,\,\sim \,\, 1:2:6  \,\,\,.
\ee
%
This pattern is obtained in a variety of schemes: $(i)$ gravity mediation, with different UV scenarios $(ii)$ gauge mediation $(iii)$ gaugino mediation \cite{Kaplan:1999ac}. We will elaborate further in the next Section.

In a previous study \cite{Arnowitt:2008bz}, we have described a series of measurements of superpartner masses at the LHC for mSUGRA models using end-point techniques \cite{Hinchliffe:1996iu}.

On the other hand, gaugino masses may be dominated entirely by one loop contributions coming from the SUGRA compensator, determined by the conformal anomaly of the effective theory at TeV scale. At some level, the flavor problem embedded in dominant tree level contributions is an indication that tree level is perhaps not the end of the story. For pure anomaly mediation, $M_a(Q)/g_a^2(Q)$ is proportional to the respective beta-function coefficients at the scale $Q$. Anomaly mediation effects are always present and become dominant over gravity mediation when the supersymmetry breaking sector is sequestered from the visible sector; however, the scheme by itself suffers from tachyonic sleptons. 

Mirage mediation is a hybrid of the mSUGRA pattern and anomaly pattern, in which the tree level contribution and the one loop contribution are both equally competitive. The quantity $M_a/g_a^2$ then depends both on the universal mSUGRA contribution, as well as the RG beta coefficients, with comparative strengths given by a parameter $\alpha$:
\be
\frac{M_a(Q)}{g_a^2(Q)} \, = \, \left(1+ \frac{\ln(M_p/m_{3/2})}{16\pi^2}g^2_{GUT}b_a \alpha \right) \frac{M_0}{g^2_{GUT}}\,\,\,,
\ee
with $M_0 \sim 1$ TeV a mass scale.

The above facts translate into the following ratio of low energy gaugino masses
\bea
{\rm Anomaly:} \,\,\,\,\,&& M_1:M_2:M_3 \,\, \sim \,\, 3.3:1:9 \,\,, \nonumber \\
{\rm Mirage:} \,\,\,\,\, && M_1:M_2:M_3 \,\, \sim \,\, \nonumber \\ 
&& (1+0.66\alpha):(2+0.2\alpha):(6-1.8\alpha) \,\,.
\eea
%
%
%

Clearly, a deviation of the measured gaugino spectrum from the mSUGRA pattern would signal the relative importance of one loop contributions (which, of course, could have a plethora of model origins). 

\subsection{Mass Measurements at the LHC}

In this paper, we take the first bottom-up steps toward the diagnosis of mediation schemes at the LHC, by choosing a representative model and attempting to reconstruct low energy masses. 

Mass reconstruction gives us the following information:

$(i)$ As we have outlined above and spell out in more detail later, mass measurements can distinguish between mediation schemes.
%
%
Of course, the choice of the UV model must be judicious - for example, reconstructing just the gaugino sector would still leave one incapable of distinguishing between the various models and schemes which give rise to the mSUGRA pattern of masses (this has also been verified by distinguishability tests \cite{Allanach:2011ya}). We thus choose a model (KKLT compactification) where the imprints of anomaly mediation are unmistakable.

One important point in this paper is that in certain regions of parameter space, the crucial information from the gaugino spectrum regarding the scheme may be gleaned at relatively lower luminosity by a judicious choice of observables. Meanwhile, the details of a particular model only become apparent as other masses are reconstructed.  

$(ii)$ The measurement of low energy masses enables us to make statements about the dark matter relic density. In models with $R-$parity invariance, the lightest superpartner (typically the lightest neutralino $\tilde{\chi_1}^0$) is a dark matter candidate, and one would like to compute the relic density $\Omega h^2$ associated with it. The direct measurement of stop ($\tilde{t}$) and stau ($\tilde{\tau}$) masses enables us to verify whether we are in regions of parameter space where stop-neutralino ($\tilde{t}-\tilde{\chi_1}^0$) and stau-neutralino ($\tilde{\tau}-\tilde{\chi_1}^0$) coannihilation effects are important. We note that the investigation of mediation schemes, the first goal described above, typically allows one to be in much larger areas of parameter space, and one need not necessarily work in coannihilation regions. However, to satisfy relic density constraints, they are crucial. 

Previously, we have undertaken studies of mass measurements in the $\tilde{\tau}-\tilde{\chi_1}^0$ coannihilation region in the context of the mSUGRA model \cite{Arnowitt:2008bz, arXiv:1008.3380}. The measurement of third generation squark masses presents its own challenges. Reconstruction of stop and sbotttom ($\tilde{b}$) is very hard in a cascade decay chain since both stops and sbottoms decay into $b$ quarks. Further, to make the situation worse, in the stop coannihilation region the stop decay produces a lower $p_T$ jet due to the proximity of the lighter stop and the lightest neutralino masses. We invoke two new observables to measure lighter stop and sbottom masses in the cascade decays.

Thus, in this paper we fulfil the following objectives: 

$(i)$ Constructing observables using different combinations of the jets, $\tau$'s and $W$'s in the final states to determine the masses of supersymmetric particles in a model where anomaly contributions are important. We construct observables in the stop-neutralino and stau-neutralino coannihilation regions. 

$(ii)$ Using the observables, we solve for the masses of the gluino, the two lighter neutralinos, the squarks (of the first two generation) and the lightest stau. We also determine the lighter stop and sbottom masses in the stop coannihilation region. 

$(iii)$ Using the neutralino and gluino masses we determine the gaugino unification scale, thereby establishing the imprint of one-loop anomaly contributions competitive with tree level contributions (mirage pattern) from experimentally measurable observables.

$(iv)$ Using the masses of the other particles in conjunction with the gauginos, we (a) determine the parameters defining a particular UV completion of the mirage pattern (KKLT compactification with visible sector on $D7$ branes) and (b) test whether we are in a coanihilation region and dark matter relic density is satisfied.

A caveat is in order. The observables we construct in the current work enable us to measure the masses of the gluino and the two lighter \textit{neutralinos}, which should be mainly gauginos for us to make statements on mediation within the structure outlined above. In the coannihilation regions, this is guaranteed, and thus we are able to simultaneously solve the relic density as well as determine the mediation scheme. It would be very interesting to probe similar issues in regions of parameter space where the Higgsino component in the neutralinos is not insignificant. However, such a study is beyond the scope of this paper.

Another comment pertains to the fact that we will be solving the model parameters from the masses. As we have stressed, our aim in this paper is to show that the determination of a given set of masses can unmistakably establish the contours of a mediation scheme (point $(iii)$ above), while it may point back to different \textit{models}. In point $(iv)$ above, we will thus not attempt a distinguishability test of our particular model (KKLT) - rather, we will fit our model parameters with the masses we solve from the observables.




The plan of the paper is as follows. In Section \ref{model}, we describe the contributions to gaugino masses, describe our choice of the UV model in more detail, and fix our benchmark points. In Section \ref{stopobservables}, we develop the kinematic observables to measure sparticle masses in the stop coannihilation region. Included in this section is the development of specific observables required for the reconstruction of stop and sbottom masses. In Section \ref{gaugino}, we give results for the gaugino mass pattern and unification scale and solve for the other masses and the full parameter set defining the UV model, along with the relic density. Moreover, we solve for the stop and sbottom masses. In Section \ref{staucoannihilation}, we perform the above analysis in the stau coannihilation region of parameter space. We end with our conclusions.

\section{Model and Benchmark Points} \label{model}


As emphasised in Section \ref{intro}, a given pattern of low energy gaugino masses accommodates diverse mediation schemes, not to mention multiple models. We mention below a catalogue of such models and schemes, outlined in \cite{Choi:2007ka}, to better situate our particular UV completion.

\subsection{Contributions to Gaugino Masses}

The various contributions to $M_a/g_a^{2}$ in $4D$ effective supergravity are
\be
\frac{M_a}{g_a^2} \,\, = \,\, \tilde{M_a}^{(0)} + \tilde{M_a}^{(1)}|_{\rm anomaly} + \tilde{M_a}^{(1)}|_{\rm other} \,\,, 
\ee
where $\tilde{M_a}^{(0)} = 1/2 F^I \partial_I f_a^{(0)}$, with $F^I$ signifying the non-zero $F-$component of a hidden sector field $X_I$ breaking supersymmetry, and $f_a$ denoting the visible sector gauge kinetic functions. The one loop anomaly term receives contributions from the conformal and Konishi anomalies, with the conformal anomaly contribution given by $\tilde{M_a}^{(1)}|_{\rm conformal} = (1/16\pi^2)b_a (F^C/C)$, with $C$ denoting the chiral compensator. We will not deal with the Konishi anomaly contribution in any detail in this paper. The other one loop contributions $\tilde{M_a}^{(1)}|_{\rm other}$ encapsulate field theoretic gauge threshold and UV-sensitive contributions like KK thresholds to the gaugino masses. We will assume that UV-sensitive contributions are subdominant (the assumption is purely because they are more model-dependent).  

The mSUGRA pattern, Eq.~\ref{mSUGRApattern}, arises in situations where $M_a/g_a^2$ is dominated by universal tree level piece $\tilde{M_a}^{(0)}$. This happens in a variety of schemes: $(i)$ gaugino mediation in higher dimensional brane models $(ii)$ gravity mediation, with different UV scenarios such as dilaton/moduli mediation in heterotic string theory or large volume compactification in type IIB tring theory. Gauge mediation also gives the mSUGRA pattern of gaugino masses, since the gaugino masses are dominated by universal one-loop gauge threshold contributions.

On the other hand, the mirage pattern in which tree level and loop level contributions are competitive occurs in gravity mediation in various UV scenarios, for example KKLT flux compactification with visible sector on $D7$ branes and explicit supersymmetry breaking by anti-$D3$ branes or spontaneous breaking by a matter sector. Mirage patterns are also obtained in  deflected anomaly mediation \cite{Rattazzi:1999qg}.

However, the important point is that deviations from the mSUGRA pattern may  signal some anomaly contribution at work. 

We will take as an example the UV completion of KKLT with visible sector on $D7$ branes and explicit supersymmetry breaking by an anti-$D3$ brane.


\subsection{KKLT with $D7$ branes: The UV Model Parameters}


In this subsection, we sketch the outlines of the UV model, only with a view to defining the model parameters we will be working with. For details, we refer to \cite{Kachru:2003aw}.

The basic elements in a KKLT-type model of string compactification are: $(i)$ background fluxes on a type IIB Calabi-Yau three-fold giving a Gukov-Vafa-Witten superpotential contribution, and $(ii)$ gaugino condensation on $D7$-branes or Euclidean $D3$ instantons giving a non-perturbative superpotential contribution. These stabilize complex structure moduli and the dilaton, as well as K\"ahler moduli, in an AdS vacuum. Supersymmetry breaking by an anti-$D3$-brane then lifts the solution to a de Sitter vacuum. The visible sector is constructed with $D7$ branes in the bulk. 

The soft masses at the GUT scale are
\begin{eqnarray} \label{softGUT}
M_a&=& M_0 +\frac{m_{3/2}}{16\pi^2}\,b_ag_a^2,
\nonumber \\
m_i^2&=& \tilde{m}_i^2-\frac{m_{3/2}}{16\pi^2}M_0\,\theta_i
-\left(\frac{m_{3/2}}{16\pi^2}\right)^2\dot{\gamma}_i
\end{eqnarray}
where $M_0$ and $\tilde{m}_i$ are pure tree level modulus contributions, given as functions of the K\"ahler modulus $T$, while other terms are one-loop anomaly contributions. $m_{3/2}$ is the gravitino mass. In the above,
\begin{eqnarray}
b_a&=&-3{\rm tr}\left(T_a^2({\rm Adj})\right)
        +\sum_i {\rm tr}\left(T^2_a(\phi_i)\right),
\nonumber \\
\gamma_i&=&2\sum_a g^2_a C^a_2(\phi_i)-\frac{1}{2}\sum_{jk}|y_{ijk}|^2,
\nonumber \\
\dot{\gamma}_i&=&8\pi^2\frac{d\gamma_i}{d\ln Q},\nonumber \\
\theta_i&=& 4\sum_a g^2_a C^a_2(\phi_i)-\sum_{jk}|y_{ijk}|^2
\frac{\tilde{A}_{ijk}}{M_0},
 \end{eqnarray}
where the quadratic Casimir $C^a_2(\phi_i)=(N^2-1)/2N$ for a fundamental representation $\phi_i$ of the gauge group $SU(N)$, $C_2^a(\phi_i)=q_i^2$ for the $U(1)$ charge $q_i$ of $\phi_i$, and $\sum_{kl}y_{ikl}y^*_{jkl}$ is assumed to be diagonal. 

To list the set of parameters that give the low energy mass spectrum, we define the ratios
\begin{equation}
\alpha\,\equiv\,\frac{m_{3/2}}{M_0\ln(M_{Pl}/m_{3/2})},\quad
1- n_i\,\equiv\, \frac{\tilde{m}_i^2}{M_0^2},
 \end{equation}
where $\alpha$ represents the anomaly to modulus mediation ratio, while $n_i$ are modular weights, that parameterize the pattern of the pure modulus mediated soft masses. 

For convenience, we will henceforth choose to rescale $\alpha$ as follows
\be \label{rescaledalpha}
\alpha_{\rm henceforth} = \frac{16\pi^2}{\ln(M_p/m_{3/2})}\frac{1}{\alpha} \,\,.
\ee

From Eq.~\ref{softGUT}, then, it is clear that the parameters defining the low energy soft masses of the model are
\begin{equation}
m_{3/2}, \, \alpha, \, n_m, \, n_H, \, {\text {tan}}\beta,
\end{equation}
where we choose $m_{3/2}$ in place of $M_0$, split the matter modular weights into a universal sfermion weight $n_m$ and a Higgs weight $n_H$. We have also replaced the Higgs mass parameters $\mu$ and $B$ by ${\rm tan}\beta$ and $M_Z$.


\subsection{Benchmark Points}



%
%


The parameter space of the KKLT model has been studied in \cite{Choi:2006im}, and we choose points that satisfy the stop-neutralino and stau-neutralino coannihilation constraints, which appear in ample parts of model parameter space. We use \darksusy\ \cite{Gondolo:2002tz} to select exact benchmark points in the above regions. 

The benchmark point for the stop coannihilation region is shown in Tables \ref{stopparametersA}, \ref{stopspectrumA}.

\begin{table}[!htp] 
\caption{Model parameters chosen at a stop coannihilation benchmark point. All masses are in GeV.}
\label{stopparametersA}
\begin{center}
\begin{tabular}{c c} \hline \hline
  Parameter      & Value   \\ \hline \hline
  $\alpha$       & 4.5   \\  
  $m_{3/2}$       & 14000   \\ 
  $n_m$       & 0.0   \\  
  $n_H$       & 0.5   \\  
  ${\rm tan}\beta$ & 30  \\  \hline \hline
\end{tabular}
\end{center}
\end{table}

\begin{table}[!htp] 
\caption{Spectrum at a stop coannihilation benchmark point. All masses are in GeV.}
\label{stopspectrumA}
\begin{center}
\begin{tabular}{cccccc} \hline \hline
  Particle      & Mass  & Particle & Mass  & Particle   & Mass   \\ \hline \hline
  $\sd_L$       & 653.13 & $\sel_L$ & 436.75 & $\neu{1}$  & 286.21  \\  
  $\sd_R$       & 635.86 & $\sel_R$ & 411.28 & $\neu{2}$  & 338.21  \\ 
  $\su_L$       & 647.91 & $\stau_1$ & 315.08 & $\neu{3}$  & 477.35  \\  
  $\su_R$       & 634.96 & $\stau_2$ & 417.70 & $\neu{4}$  & 502.68  \\  
  $\tilde{b}_1$ & 520.46 &           &       & $\chpm{1}$ & 337.32  \\  
  $\tilde{b}_2$ & 596.25 &           &       & $\chpm{2}$ & 500.41  \\
  $\st_1$       & 338.55 &           &       & $\sg$      & 649.78 \\
  $\st_2$       & 616.22	&           &       &            &        \\\hline \hline 
\end{tabular}
\end{center}
\end{table}

The benchmark point for the stau coannihilation region is shown in Table \ref{Staubenchmarkpoint}.

\begin{table}[!htp] 
\caption{Model parameters chosen at a stau coannihilation benchmark point. All masses are in GeV.}
\label{Staubenchmarkpoint}
\begin{center}
\begin{tabular}{c c} \hline \hline
  Parameter      & Value   \\ \hline \hline
  $\alpha$       & 7.5   \\  
  $m_{3/2}$       & 10000   \\ 
  $n_m$       & 0.5   \\  
  $n_H$       & 1.0   \\  
  ${\rm tan}\beta$ & 30  \\  \hline \hline
\end{tabular}
\end{center}
\end{table}

The spectrum at the stau coannihilation benchmark point is shown in Table \ref{Stauspectrum}.

\begin{table}[!htp] 
\caption{Spectrum at a stau coannihilation benchmark point. All masses are in GeV.}
\label{Stauspectrum}
\begin{center}
\begin{tabular}{cc cc cc } \hline \hline
  Particle      & Mass  & Particle & Mass  & Particle   & Mass   \\ \hline \hline
  $\sd_L$       & 845.49 & $\sel_L$ & 426.91 & $\neu{1}$  & 284.17  \\  
  $\sd_R$       & 813.52 & $\sel_R$ & 367.70 & $\neu{2}$  & 389.17  \\ 
  $\su_L$       & 841.39 & $\stau_1$ & 309.75 & $\neu{3}$  & 548.88  \\  
  $\su_R$       & 815.27 & $\stau_2$ & 425.68 & $\neu{4}$  & 569.04  \\  
  $\tilde{b}_1$ & 735.87 &           &       & $\chpm{1}$ & 389.32  \\  
  $\tilde{b}_2$ & 791.30 &           &       & $\chpm{2}$ & 568.10  \\
  $\st_1$       & 600.23 &           &       & $\sg$      & 897.55 \\
  $\st_2$       & 810.20	&           &       &            &        \\\hline \hline
\end{tabular}
\end{center}
\end{table}

Note that our methods are valid in general, and the above benchmark points will be explored as an illustration. In particular, we will also study benchmark points with higher gluino mass, preferred by current LHC data, in Sections \ref{gauginounifforstop} and \ref{gauginounifforstau}, where we will show that we need larger luminosity to establish the same set of observables.

Since ${\rm tan} \beta$ is on the large side for the benchmark points, the lighter stau mass is between the lightest and next to lightest neutralinos. The lightest neutralino is mostly Bino and the next to lightest is mostly Wino. The Higgsino components are negligible since the dark matter relic density is satisfied by the coannihilation mechanism.



%
%





\section{Kinematic Observables in the Stop Coannihilation Region}\label{stopobservables}


In this section, we present the measurement of physical observables that will be used to solve for the gaugino masses, as well as the masses of the $\tilde{\tau}$ and $\tilde{q}$. Moreover, we construct observables that will be needed to solve for the $\st$ and $\tilde{b}$ masses. In the next section, we will present the mass and model parameter solutions, and also present gaugino unification. We will also give results for a benchmark point with heavier gluino there.



The mass spectrum of the model is determined using \isasugra\ \cite{isajet}. The spectrum is then fed to \pythia\ \cite{pythia}, which generates the Monte Carlo hard scattering events and hadron cascade. These events are passed to the detector simulator \pgs \ \cite{pgs}.







\subsection{$2$ Jets + $2 \tau$ + $\met$}


At the LHC, the main production processes for this model are $\tilde g\tilde g$, $\tilde g \tilde q$ and $\tilde q \tilde q$.

The relevant decay chains are:
\be
\tilde g \longrightarrow \overline{q}_L \tilde q_L \nonumber
\ee
and
\be \label{decay1}
{\tilde {q}}_{L} \longrightarrow \neu{2}q \longrightarrow {\tilde {\tau }}_{1}^{\pm } \tau ^{\mp } q \longrightarrow \neu{1} \tau ^{\pm } \tau ^{\mp } q \,\,\,.
\ee
There is a high $p_T$ jet, missing transverse energy, and a pair of oppositely charged $\tau$ leptons.

The following cuts were set to select events for this signal: 

$(i)$ Missing transverse energy $\met \geq 180$ GeV;

$(ii)$ $p_{T,jet1}+p_{T,jet2} + \met\geqslant 600$ GeV;

$(iii)$ Leading jet cuts: At least two jets should be present, each with $p_T \geq 200$ GeV in $|\eta| \leq 2.5$. The jets are both required to be non $b$-tagged jets, and the event is discarded when either of them is tagged as a $b$ jet;

$(iv)$ Soft jet cuts: Any jet with $p_T \geq 30$ GeV in $|\eta| \leq 2.5$ is accepted in the analysis;

$(v)$ $\tau$ cuts \cite{CMStau}: At least two $\tau$ leptons, with visible $p_T \geq 15$ GeV in $|\eta| \leq 2.5$;

Various kinematic observables can be constructed in this signal. Of them, we choose four that give the most precise endpoint measurements with $50$ fb$^{-1}$: $M^{\rm end}_{j\tau \tau }$, $M^{\rm end}_{\tau \tau }$, and the $p_T$ distributions of the higher and lower energy $\tau'$s.  

The fifth observable (required to solve the five model parameters) is the peak of the $M_{\rm eff}$ distribution.


\subsubsection{$M^{\rm end}_{\tau \tau }$}


The main challenge in measuring the invariant di-tau mass is the background created by uncorrelated $\tau$ pairs. From the decay chain in question, it is clear that similarly charged $\tau$'s are definitely uncorrelated and hence model this background quite well. Thus, we perform the opposite-sign (OS) minus like-sign (LS) subtraction on $\tau$'s to obtain the signal.

\begin{figure}[!htp]
\centering
\includegraphics[width=3.5in]{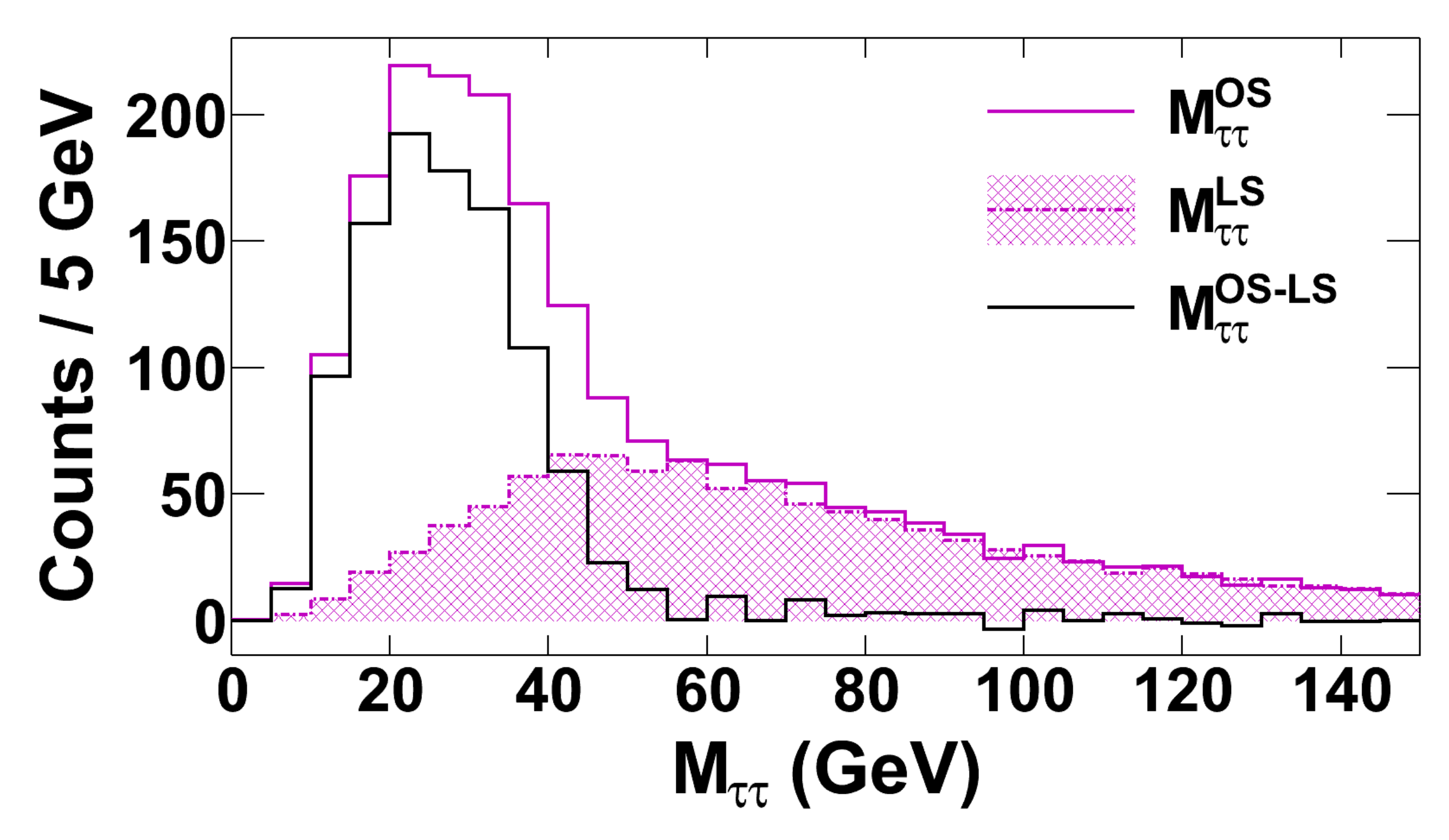}
\caption{Distribution of $M_{\tau \tau }$ at a stop coannihilation benchmark point. The pink (grey) histogram is composed of OS $\tau$ pairs. The filled dot-dashed pink (grey) histogram is composed of LS $\tau$ pairs and is normalised to the shape of the long tail of the the pink (grey) OS histogram. The OS$-$LS subtraction produces the black subtracted histogram. This subtracted histogram is then fitted with a straight line to find the endpoint of the distribution. The result for the endpoint is $49.71 \pm 0.2 (\rm Stat.)$ GeV. The luminosity is $50$ fb$^{-1}$. }
\label{StopMtt}
\end{figure}

Figure \ref{StopMtt} shows the histogram graph we obtained for the $M_{\tau \tau }$ distribution. The luminosity is $50$ fb$^{-1}$, and we can see that a clear end-point is obtained.

\subsubsection{$M^{\rm end}_{j\tau \tau }$}


The $\tau$ leptons from each event are sorted into OS and LS pairs. Then, each pair is combined with the leading jets from the same event, and the invariant mass distribution $M_{j\tau \tau }^{\rm same}$ is obtained. At this stage the Bi-Event Subtraction Technique (BEST) \cite{Dutta:2011gs} is initiated, by first combining the $\tau$ pair with leading jets from a separate event (which we also call a bi-event), to form the distribution $M_{j\tau \tau }^{\rm bi-event}$. Since jets from a separate event are kinematically uncorrelated with the $\tau$ pair from the current event, this bi-event distribution models the jet background very well. 

We note that the greater the number of different-event jets included in the bi-event distribution, the better the modelled background becomes. We include sufficient number of different-event jets to ensure a smooth background distribution.

\begin{figure}[!htp]
\centering
\includegraphics[width=3.5in]{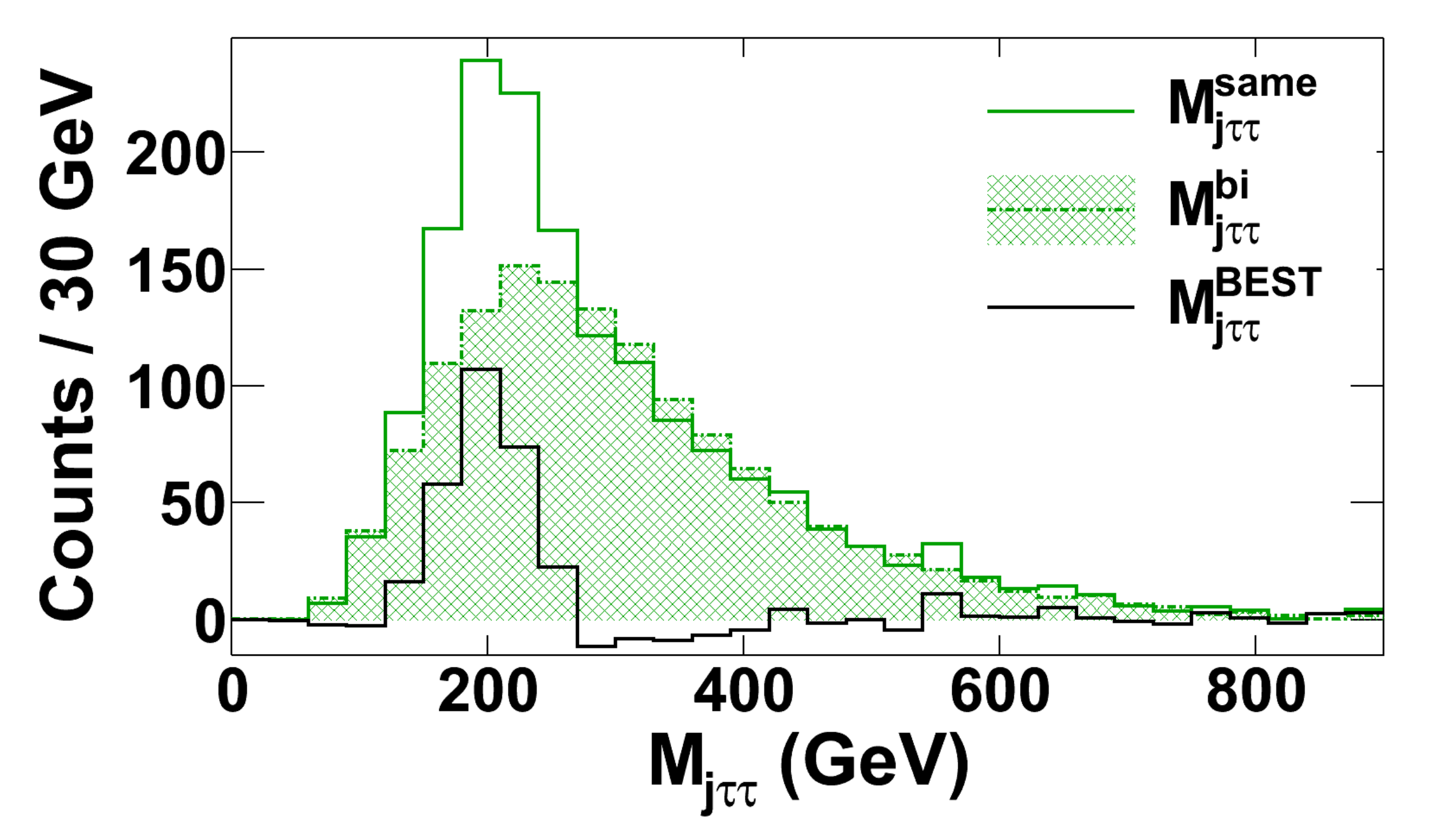}
\caption{Distribution of $M_{j\tau \tau }$ at a stop coannihilation benchmark point. The green (grey) same-event histogram is constructed by combining each OS$-$LS $\tau$ pair with a leading jet from the same event. The filled dot-dashed green (grey) bi-event histogram is obtained by combining the OS$-$LS $\tau$ pair with a leading jet from a different event and is normalised to the long tail of the same-event histogram. The same-event minus bi-event subtraction (BEST) produces the black subtracted histogram. The subtracted histogram is fitted with a straight line to obtain the endpoint. The result for the endpoint is  $269.09 \pm 3.18 (\rm Stat.)$ GeV. The luminosity is $50$ fb$^{-1}$.}
\label{StopMjtt}
\end{figure}

After this, two subtractions are performed to obtain the final signal: $(i)$ The OS$-$LS subtraction, which gets rid of uncorrelated $\tau$ pairs and $(ii)$ The BEST subtraction, which gets rid of uncorrelated jets in the background. In Figure \ref{StopMjtt}, we show the $M_{j\tau\tau}$ distribution where we find a very clear end point after the BEST subtraction.

\subsubsection{Slope of $p_T$ of $\tau$}


The mass of the $\tilde{\tau}$ lies between the two lightest neutralinos, and its exact location approximately determines the $p_T$ of the $\tau$'s
in the ditau pair. The slope of the visible $p_T$ distribution of the lower energy visible $\tau$ from the di-tau pair (which we will denote by $slope(p_T)$) is generally a good observable, that carries information about the masses of $\tilde{\tau}$ and $\neu{1}$ (in the case that the lower energy $\tau$ is from $\tilde{\tau} \longrightarrow \neu{1} + \tau$). If the stau mass is very close to the neutralino mass (in the case of stau-neutralino coannihilation) we have a low energy $\tau$ and the slope of the $p_T$ distribution is a good observable. 

Conversely, if the visible $p_T$ of a $\tau$ is large, the slope of the $p_T$ distribution is not a good observable. This is because the slope in the case of a higher energy $\tau$ doesn't show enough variation. In such cases, we use the kinematic information of the high energy $\tau$ by combining it with a leading jet to construct $M_{j\tau}$. We will use the endpoint of the $M_{j\tau}$ distribution as an observable in the stau-neutralino coannihilation case, where one of the $\tau$'s has large $p_T$ compared to the other.

If the stau is somewhat in between the two neutralinos, the slopes of the  $p_T$ distributions of \textit{both} taus become good observables (provided, of course, that neither $p_T$ is too large) as happens in this case. In such a scenario, we get two observables depending on two neutralinos and the stau mass without involving any jet. It is convenient to define the new observables $p_{T,{\rm AM}}$ and $p_{T,{\rm diff}}$, which are the arithmetic mean and difference of the slopes of the higher and lower $p_T$ $\tau'$s. These variables can typically be measured down to lower luminosity than $M_{j\tau}$, since they do not involve the subtractions required for observables involving jets, as we describe below.

To obtain the distributions, first the OS$-$LS subtraction is performed. Then, we take the logarithm of the $p_{T,{\rm high}}$ and $p_{T,{\rm low}}$ distributions for the higher and lower energy $\tau$ respectively. The histograms are fitted using a linear function to find the slopes. Finally, the slopes are combined into the arithmetic mean $p_{T,{\rm AM}}$ and the difference $p_{T,{\rm diff}}$:
\bea
p_{T,{\rm AM}}  &=& \frac{1}{2}\left(slope(p_{T,{\rm high}}) + slope(p_{T,{\rm low}})\right) \nonumber \\
p_{T,{\rm diff}} &=& \frac{1}{2}\left(slope(p_{T,{\rm high}}) - slope(p_{T,{\rm low}})\right) \,\,\,.
\eea
Note that these observables, formed by combining the $p_T$ information of the two $\tau$'s, are particularly necessary in the case when they are close in mass, since we do not know if a given $\tau$ is originating from the $\tilde{\tau}$ decay or the $\neu{2}$ decay.

\begin{figure}[!htp] 
\centering
\includegraphics[width=3.5in]{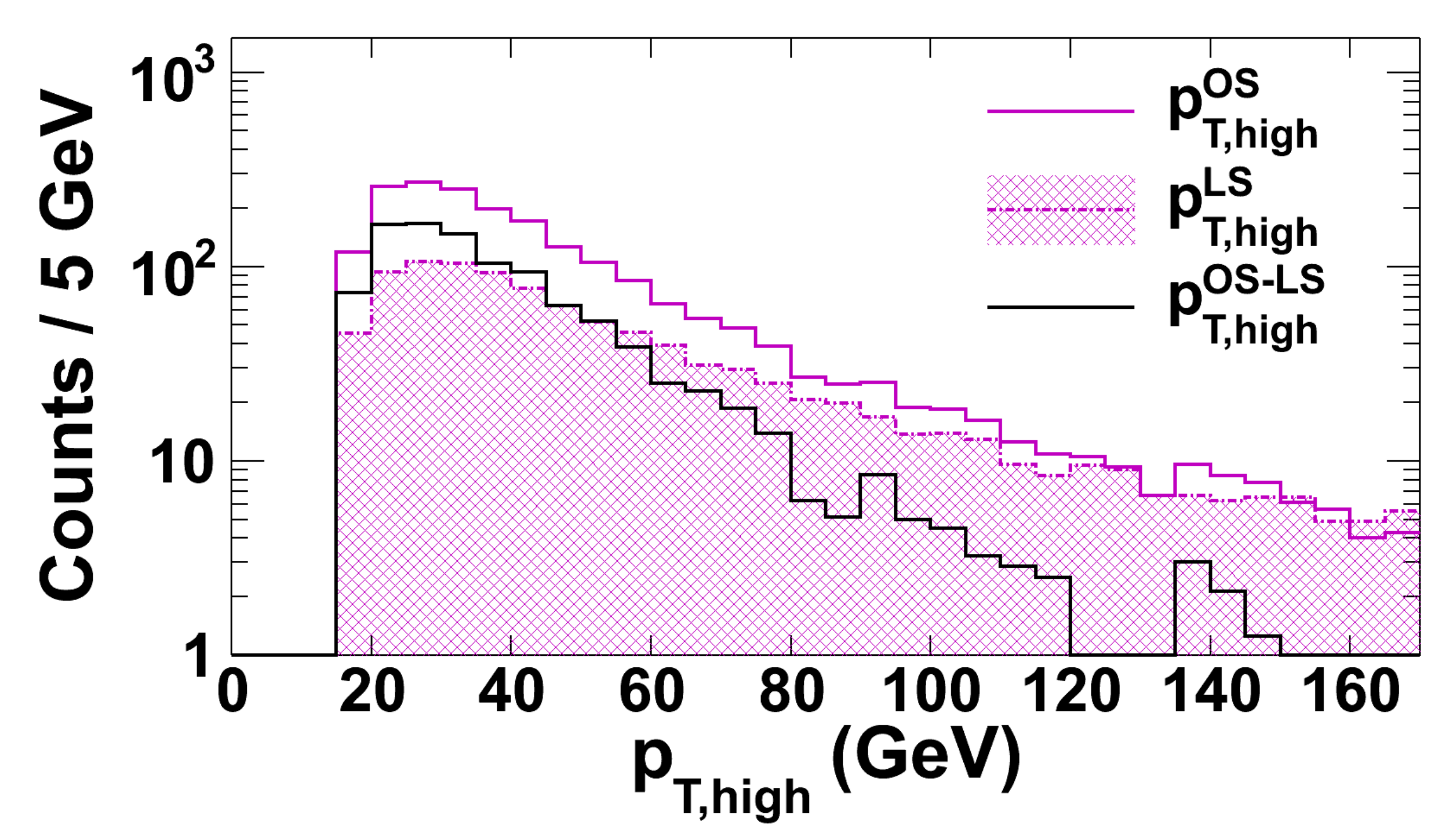}
\caption{Distribution of  $p_{T,{\rm high}}$ at a stop coannihilation benchmark point. The pink (grey) histogram is composed of OS $\tau$ pairs. The filled dot-dashed pink (grey) histogram is composed of LS $\tau$ pairs and is normalised to the shape of the long tail of the OS histogram. The OS$-$LS subtraction produces the black subtracted histogram. This subtracted histogram is then fitted with a straight line to find the slope of the distribution. The value of the $slope(p_{T,{\rm high}})$ observable is $-0.0522 \pm 0.0021 (\rm Stat.)$. The luminosity is $50$ fb$^{-1}$.}
\label{Stopptmax}
\end{figure}

\begin{figure}[!htp]
\centering
\includegraphics[width=3.5in]{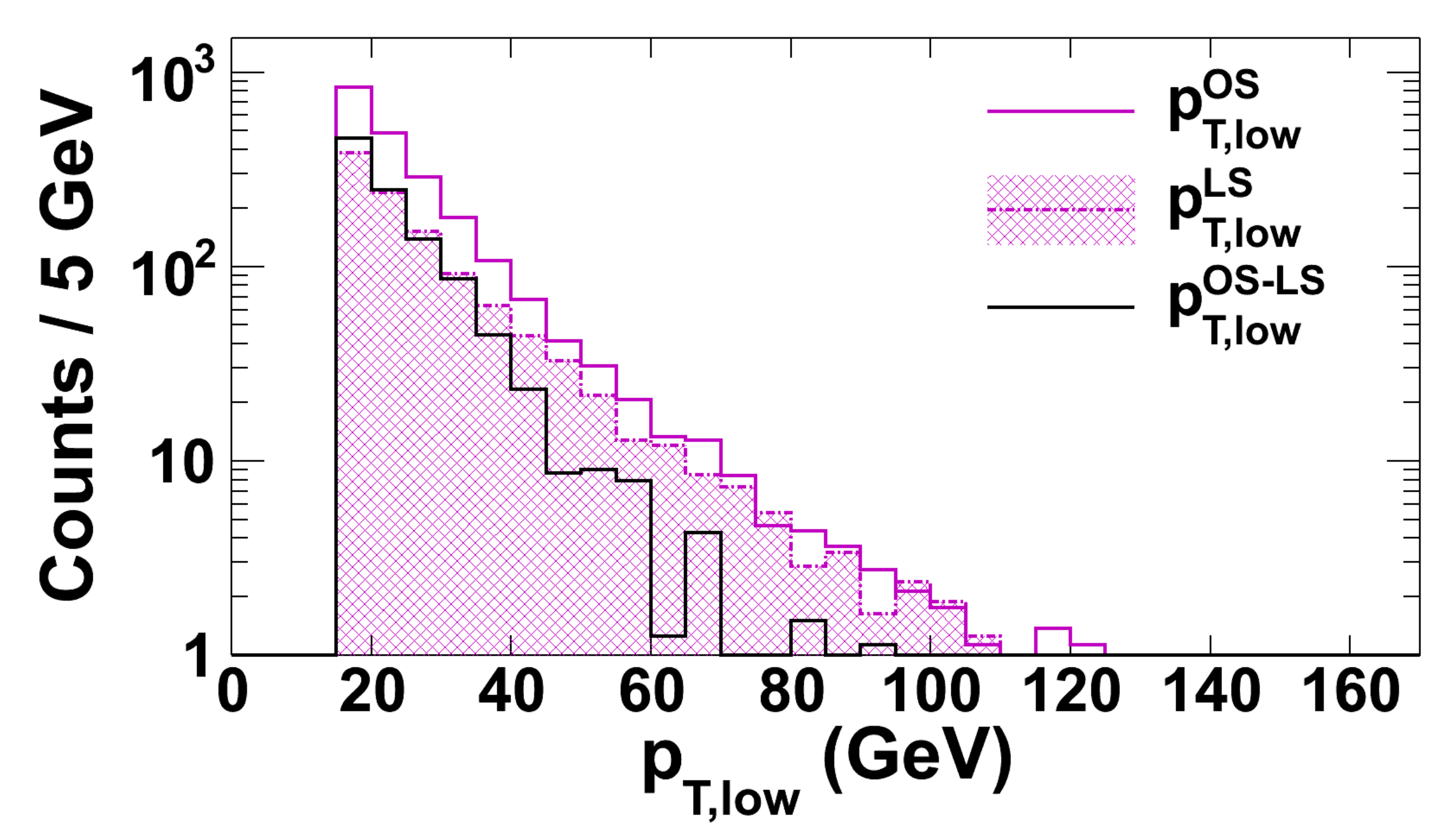}
\caption{Distribution of  $p_{T,{\rm low}}$ at a stop coannihilation benchmark point. The pink (grey) histogram is composed of OS $\tau$ pairs. The filled dot-dashed pink (grey) histogram is composed of LS $\tau$ pairs and is normalised to the shape of the tail of the OS histogram. The OS$-$LS subtraction produces the black subtracted histogram. This subtracted histogram is then fitted with a straight line to find the slope of the distribution. The value of the $slope(p_{T,{\rm low}})$ observable is $-0.1178 \pm 0.0040 (\rm Stat.)$. The luminosity is $50$ fb$^{-1}$.}
\label{Stopptmin}
\end{figure}

In Figure \ref{Stopptmax} and Figure \ref{Stopptmin}, we show the $p_T$ distributions of two $\tau$'s present in the sample.

\subsection{$4$ jets + $\met$}


The $M_{\rm eff}$ distribution is formed by combining the $p_T$ of the first four leading jets and the missing energy
\be \label{3.10}
M_{\rm eff}=p_{T,jet1}+p_{T,jet2}+p_{T,jet3}+p_{T,jet4}+\met
\ee
These jets effectively come from the gluino and squark decays.

The effective mass $M_{\rm eff}$ is constructed from the 4jets + $\met$ sample, with the following cuts

$(i)$ Number of jets: At least four jets in the event;

$(ii)$ Leading jet cuts: The first two leading jets each have $p_T \geq 200$ GeV in $|\eta| \leq 2.5$;

$(iii)$ Soft jet cuts: Jets with $p_T \geq 30$ GeV in $|\eta| \leq 2.5$ are accepted in the analysis;

$(iv)$ The jets are not $b$ tagged;

$(v)$ $\met \geq 180$ GeV;

$(vi)$ $p_{T,jet1} + p_{T,jet2} + \met \,\geqslant \, 600$ GeV;

$(vii)$ No $e'$s and $\mu'$s with  $p_T \geq 15$ GeV;

$(viii)$ Transverse sphericity $S_T \leq 0.2$ .

\begin{figure}[!htp]
\centering
\includegraphics[width=3.5in]{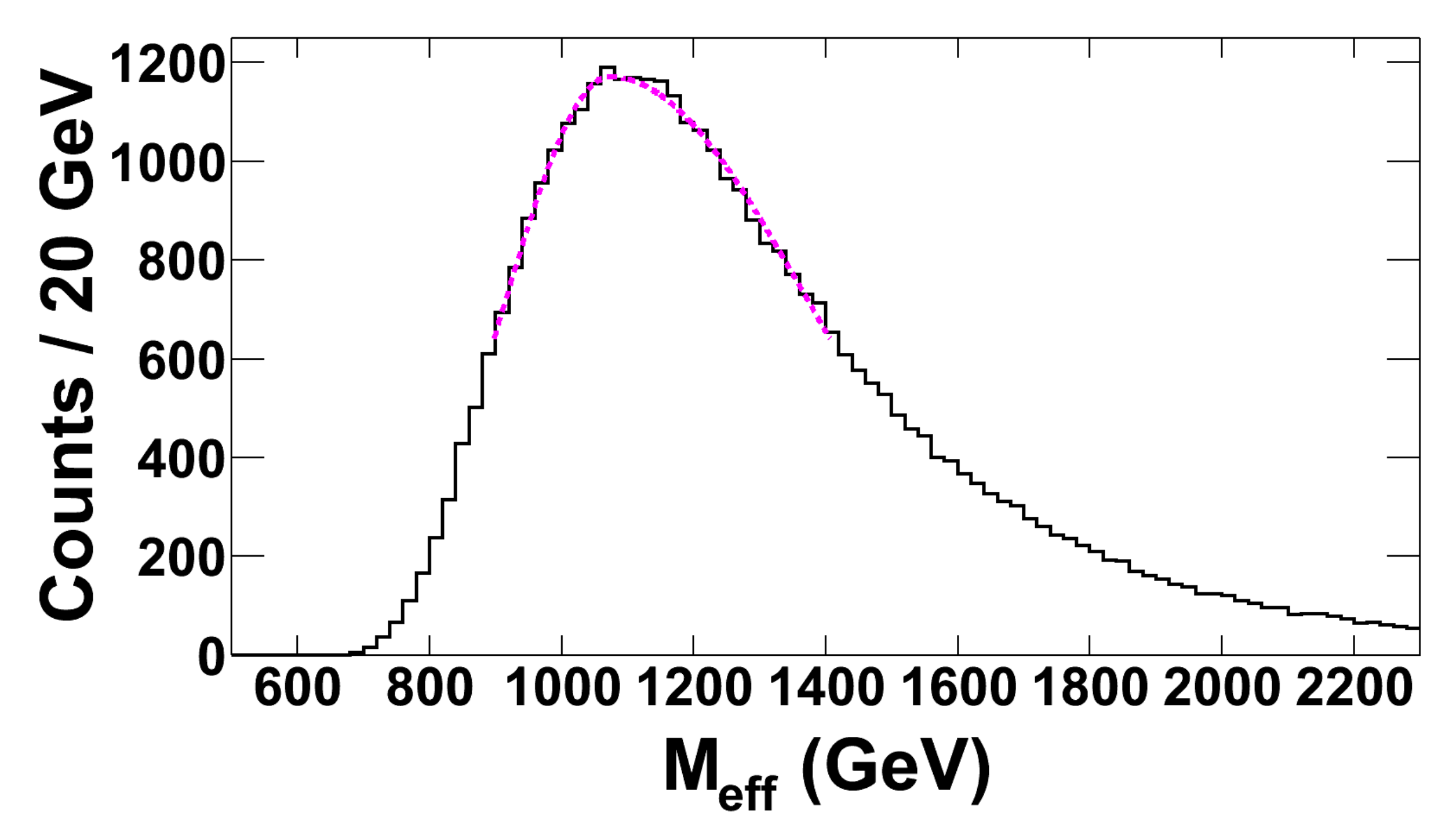}
\caption{Distribution of $M_{\rm eff}$ at a stop coannihilation benchmark point. The distribution is fitted with a Gaussian function to find the peak. The value of the  $M^{\rm peak}_{\rm eff}$  observable is  $1073.01 \pm 8.72 (\rm Stat.)$ GeV. The luminosity is $50$ fb$^{-1}$.}
\label{Stopmeff}
\end{figure}

In Figure \ref{Stopmeff}, we show the $M_{\rm eff}$ distribution at the benchmark point. The peak of this distribution shows a well defined peak for $50$ fb$^{-1}$ luminosity.

In Table \ref{Stopobservables}, we show the end points of $M_{\tau\tau}$ and $M_{j\tau\tau}$ distributions, peak of the $M_{\rm eff}$ distribution, and the slopes of the $p_T$ distributions for the taus for our benchmark point. The statistical uncertainties range between $0.4\% - 3.8\%$.

\begin{table}[!htp] 
\caption{Kinematic observables at $50$ fb$^{-1}$ for a stop coannihilation benchmark point. All masses are in GeV.}
\label{Stopobservables}
\begin{center}
\begin{tabular}{c c c c } \hline \hline
  Observable                        & Value       & $50$ fb$^{-1}$ Stat.        & $100$ fb$^{-1}$ Stat.          \\ \hline \hline
  $M^{\rm end}_{\tau \tau }$                  & $49.71$     & $\pm 0.20$                & $\pm 0.14$                             \\  
  $M^{\rm end}_{j\tau \tau }$                 & $269.09$    & $\pm 3.18$                & $\pm 2.25$                           \\ 
  $slope(p_{T,{\rm high}})$               & $-0.0522$   & $\pm 0.0021$              & $\pm 0.0014$                      \\  
  $slope(p_{T,{\rm low}})$                & $-0.1178$   & $\pm 0.0040$              & $\pm 0.0028$                        \\  
  $M^{\rm peak}_{\rm eff}$                         & $1073.01$   & $\pm 8.72$                & $\pm 6.17$                                 \\\hline  \hline 
\end{tabular}
\end{center}
\end{table}


\subsection{Observables for the Determination of Third Generation Squark Masses: $M^{\rm end}_{bW}$ and $M^{\rm end}_{jW}$}\label{thirdgeneration}


To probe the third generation squark masses we need to involve $b$ quarks. The relevant decay chain associated with the dominant production process for the reconstruction of third generation squarks in the stop coannihilation region is
\be
\tilde{g} \longrightarrow \tilde{b} + b \longrightarrow \tilde{t} + W + b \longrightarrow \neu{1} + c + b + W.
\ee

These signals are characterized by high $p_T$ jets accompanied by a $W$ and $\met$.

We will be constructing the distributions $M_{bW}$ and $M_{jW}$.

The cuts for the analysis are

$(i)$ $\met \geq 180$ GeV;

$(ii)$ Number of jets: $N_{\rm jet} \geq 4$;

$(iii)$ Leading jet cuts: The first two leading jets each have $p_T \geq 200$ GeV in $|\eta| \leq 2.5$. They could be gluon, light-flavor, or $b$ jets;

$(iv)$ Soft jet cuts: Any jets with visible $p_T \geq 30$ GeV in $|\eta| \leq 2.5$ are accepted in the analysis. This includes $b$-tagged jets;

$(v)$ $p_{T,jet1}+p_{T,jet2} + \met\geqslant 600$ GeV;

$(vi)$ For $M_{bW}$, at least one tight $b$ jet is required.

The first step in the analysis is the reconstruction of the $W$ boson. The $W$ appears in the detector as two jets whose invariant mass falls in the $W$ mass window ($65$ GeV $\leq M_{jj} \leq 90$ GeV). We thus choose soft jet pairs (from the third leading jet and below) which are not $b$-tagged, with $0.4 \leq \Delta R \leq 1.5$. The jets are put into two categories: those which are manifestly in the $W$ window, and those that fall within the sideband window ($40$ GeV $\leq M_{jj} \leq 55$ GeV or $100$ GeV $\leq M_{jj} \leq 115$ GeV). BEST is then performed for the two categories, to get rid of uncorrelated jet background. After this, the sideband subtraction is performed to obtain the $W$ mass.

Once the $W$ is reconstructed, it is paired up with jets to form the $M_{bW}$ and $M_{jW}$ distributions.

\begin{figure}[!htp] 
\centering
\includegraphics[width=3.5in]{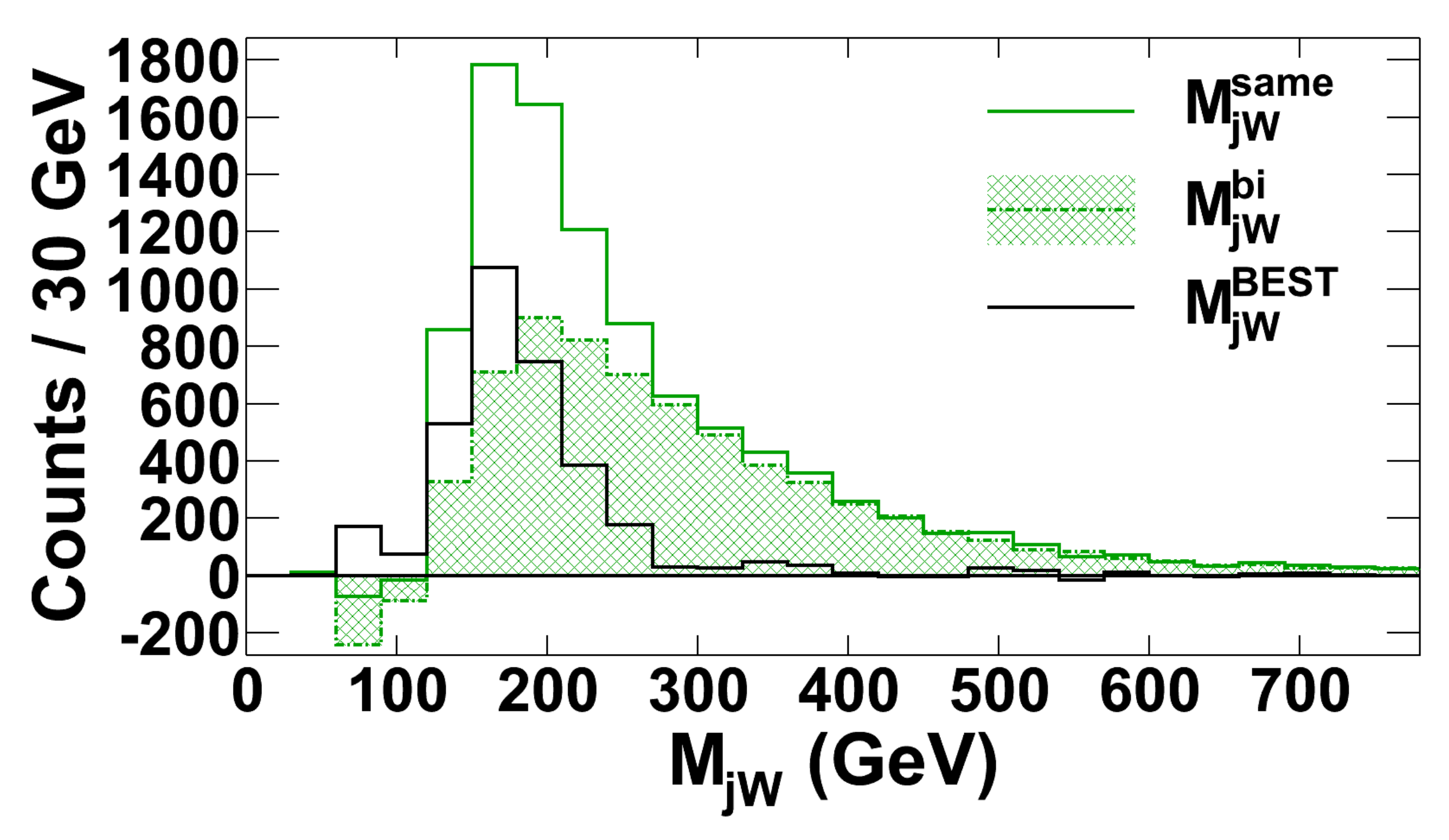}
\caption{Distribution of $M_{jW}$ at a stop coannihilation benchmark point. $W$ is first reconstructed with two jets whose invariant mass falls in the $W$ window. The reconstructed $W$ is then combined with a non $b$-tagged soft jet of rank three or lower from the same event, to produce the same-event blue (grey) histogram. The $W$ is combined with a soft jet from a different event to produce the bi-event filled dot-dashed blue (grey) histogram, which is normalised to the shape of the long tail of the same-event histogram. The same-event minus bi-event subtraction (BEST) produces the black subtracted histogram. The subtracted histogram is fitted with a straight line to obtain the endpoint. The result for the endpoint is  $287.55 \pm 0.74 (\rm Stat.)$ GeV. The luminosity is $50$ fb$^{-1}$.}
\label{StopMjw}
\end{figure}

\begin{figure}[!htp]
\centering
\includegraphics[width=3.5in]{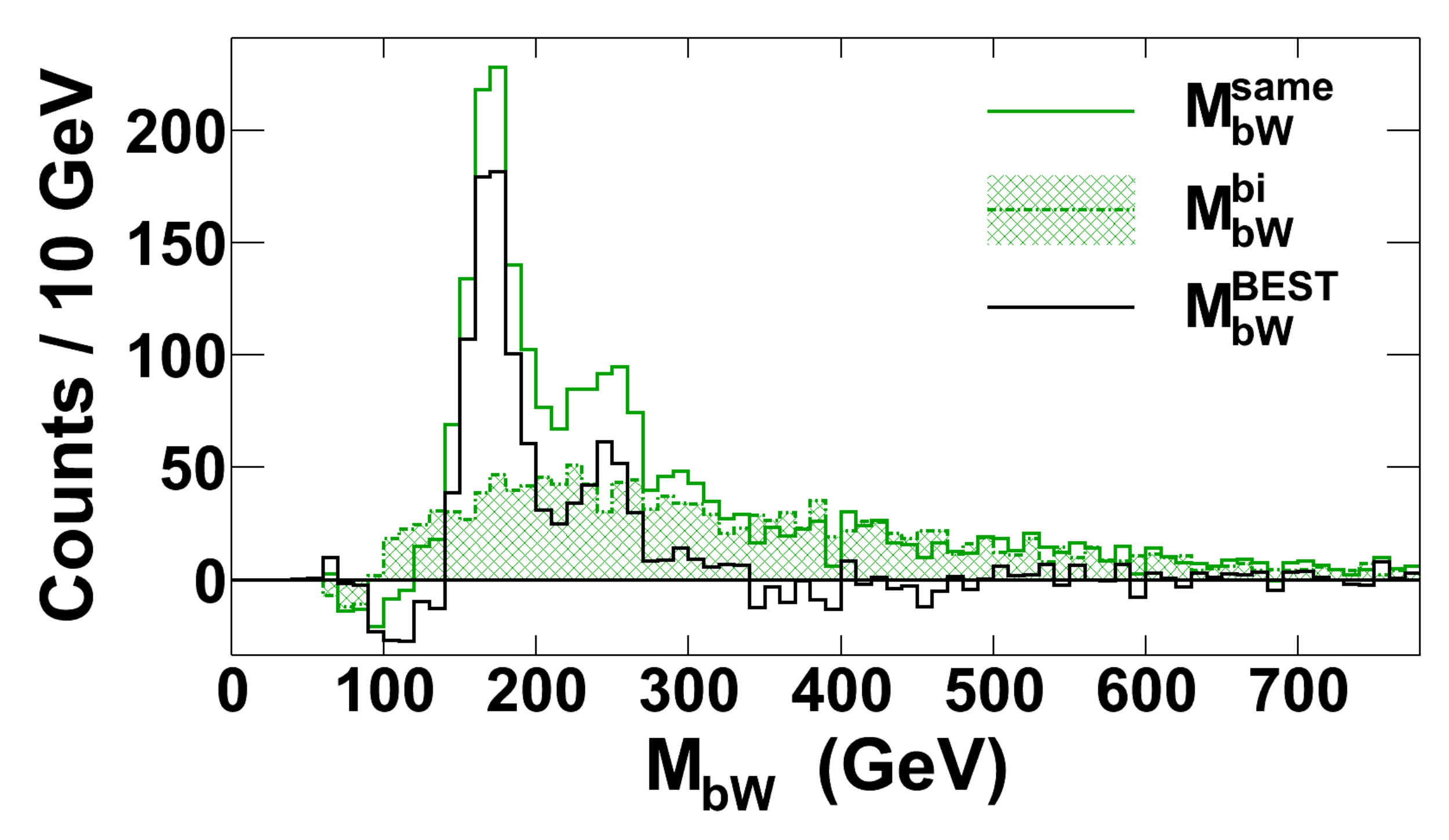}
\caption{Distribution of $M_{bW}$ at a stop coannihilation benchmark point. $W$ is first reconstructed with two jets whose invariant mass falls in the $W$ window. The reconstructed $W$ is then combined with a $b$ jet of any rank from the current event, to produce the same-event pink (grey) histogram. Events with $M_{bW} \leq 200$ GeV are discarded to remove the top peak. The $W$ is combined with a $b$ jet from a different event to produce the bi-event filled dot-dashed pink histogram, which is normalised to the shape of the long tail of the same-event histogram. The same-event minus bi-event subtraction (BEST) produces the black subtracted histogram. The subtracted histogram is fitted with a straight line to obtain the endpoint. The result for the endpoint is  $325.67 \pm 4.50 (\rm Stat.)$ GeV.  The luminosity is $50$ fb$^{-1}$.}
\label{StopMbw}
\end{figure}

For the $M_{jW}$ distribution, we pair the $W$ with a non $b$-tagged soft jet, whose rank is three or lower. This is because we are in the stop coannihilation region. In Figure \ref{StopMjw}, we show the $M_{jW}$ distribution at the benchmark point, finding a well defined end-point for $50$ fb$^{-1}$ luminosity. 

For the $M_{bW}$ distribution, we pair the $W$ with a $b$ jet of \textit{any} rank from the current event. Note that the relevant $b$ jet required to construct this observable need not be a leading jet; in fact, the leading jet will typically be non $b$-tagged. After pairing the $W$ with the $b$ jet, we do a further BEST to get rid of uncorrelated $b$ jets. This gives the final signal for $M_{bW}$. However, the $b+W$ signal shows the presence of the unwanted top peak, which comes from $t \longrightarrow b+W$. The top window is removed from the final signal, by discarding events with $M_{bW} \leq 200$ GeV. In Figure \ref{StopMbw}, we show the $M_{bW}$ distribution obtained at the benchmark point, finding the end-point for $50$ fb$^{-1}$ luminosity.

In Table \ref{Stopthirdgenobservables}, we show the endpoint values obtained from these distributions. The statistical uncertainties range between $0.2\% - 1.4\%$. The statistical uncertainty is larger for $M_{bW}$ due to the $b$ jet.

\begin{table}[!htp] 
\caption{Kinematical observables $M^{\rm end}_{jW}$ and $M^{\rm end}_{bW}$ at $50$ fb$^{-1}$ for a stop coannihilation benchmark point. All masses are in GeV.}
\label{Stopthirdgenobservables}
\begin{center}
\begin{tabular}{c c c c } \hline \hline
  Observable                    & Value       & $50$ fb$^{-1}$ Stat.        & $100$ fb$^{-1}$ Stat.          \\ \hline \hline
  $M^{\rm end}_{jW}$                & $287.55$    & $0.74$                      & $0.52$                          \\
  $M^{\rm end}_{bW}$                & $325.67$    & $4.50$                      & $3.18$                         \\ \hline \hline
\end{tabular}
\end{center}
\end{table}



\section{Determination of Particle Masses, Gaugino Unification Scale, and UV Model Parameters in the Stop Coannihilation Region}\label{gaugino}

The observables measured in Section \ref{stopobservables} are used to determine the masses of the gluino, $\neu{1}$, $\neu{2}$, $\tilde{\tau}$, and $\tilde{q}$, which lead to the determination of model parameters $(\alpha, m_{3/2}, {\rm tan} \beta, n_m, n_H)$. We will also determine the $\tilde{b}$ and $\tilde{t}$ masses later in this section.

Theoretically, the functional dependences of the observables in Section \ref{stopobservables} are expected to be: 
\bea \label{theoryfunctions}
&& M^{\rm end}_{\tau \tau } \, = \,M^{\rm end}_{\tau \tau }(m_{\neu{2}}, m_{\tilde{\tau}}, m_{\neu{1}}) \nonumber \\
&& M^{\rm end}_{j\tau \tau } \, = \, M^{\rm end}_{j\tau \tau }(m_{\tilde{q_L}},m_{\neu{2}}, m_{\neu{1}}) \nonumber \\
&& slope(p_{T,{\rm AM}}) \, = \, p_{T,{\rm AM}}(m_{\neu{2}}, m_{\neu{1}}) \nonumber \\
&& slope(p_{T,{\rm diff}}) \, = \, p_{T,{\rm diff}}(m_{\tilde{\tau}}, m_{\neu{2}}, m_{\neu{1}}) \nonumber \\
&& M^{\rm peak}_{\rm eff} \, = \, M^{\rm peak}_{\rm eff}(m_{\tilde{q_L}}, m_{\tilde{g}}) \nonumber \\
\eea

The masses are varied independently around the benchmark point, and the full simulation of the collider experiment and determination of the observables is performed each time. Thus, the functional dependence of the observables on the masses is determined, along with their uncertainties. 

This set of equations can be used to solve for the masses. In general, the dependence of the observables on the masses is expected to be non-linear, corresponding to multiple solutions of the masses. However, near the benchmark point, we found that for most masses, linear functions fitted the dependence quite well. We obtained two solutions only for the stau mass, and one was quite far from the benchmark point and thus rejected. We will have more to say on these issues in Section \ref{staumassgaugunifsection}.




In Table \ref{Stopmasssolutions}, we show the solution to the $\sg, \tilde{q}_L, \tilde{\tau}, \neu{2},$ and $\neu{1}$ masses for $50$ fb$^{-1}$ luminosity. We also show the statistical uncertainties, which range between $1\% - 6.6\%$.

\begin{table}[!htp] 
\caption{Solution to masses at $50$ fb$^{-1}$ for a stop coannihilation benchmark point. All masses are in GeV.}
\label{Stopmasssolutions}
\begin{center}
\begin{tabular}{c c c c } \hline \hline
  Particle      & Mass  & $50$ fb$^{-1}$ Stat. & $100$ fb$^{-1}$ Stat.    \\ \hline \hline
  $\sg$       & 646 & $-$14,+19 & $-$11,+14    \\  
  $\tilde{q}_L$       & 638 & $-$34,+42 & $-$23,+39   \\ 
  $\tilde{\tau}$       & 318 & $-$3,+3 & $-$3,+3   \\  
  $\neu{2}$       & 333 & $-$7,+11 & $-$6,+8   \\  
  $\neu{1}$ & 276 & $-$8,+13 & $-$7,+10     \\\hline \hline  
\end{tabular}
\end{center}
\end{table}









\subsection{Determination of UV Model Parameters and Relic Density}\label{modelparameters}


Having determined the masses of $\tilde{g}, \neu{1}, \neu{2}, \tilde{\tau},$ and $\tilde{q}$, we can use these to numerically solve for the parameters of the full model. As we have stressed in Section \ref{intro}, the point of this work is to show that the contours of the mediation scheme (here, mirage pattern) show up unmistakably from the determination of certain masses. We will demonstrate this in Section \ref{gauginounifforstop}, where we will show the gaugino unification scale. However, a set of masses may point back at different \textit{models}, and in particular, we will not attempt to perform a distinguishability study of our particular model. When we solve the model parameters for our particular example (KKLT), our methods will reflect this - we will start from the \textit{model parameters}, fit with the spectrum, and zero in on the \textit{masses} we solved. 

We use the Nelder Mead method, a commonly used nonlinear optimization technique, to do so. An initial simplex is chosen on the parameter space $\{m_{3/2}, \, \alpha, \, n_m, \, n_H, \, {\text {tan}}\beta,\}$ and \isajet\ \cite{isajet} is used to find the corresponding mass. The scan over parameter space proceeds until the masses calculated by \isajet\ are close to the solved values.

Table \ref{Stopmodelparametersolution} shows the corresponding model parameters we determined from  $m_{\tilde {g}}$, $m_{{\tilde {q}}_{L}}$, $m_{\neu{2}}$, $m_{{\tilde {\tau }}_{1}}$ and $m_{\neu{1}}$ for $50$ fb$^{-1}$ and $100$ fb$^{-1}$ luminosities. At $50$ fb$^{-1}$, the statistical uncertainties range between $5.6\%-17\%$.

\begin{table}[!htp] 
\caption{Solution to model parameters at a stop coannihilation benchmark point at $50$ fb$^{-1}$. Masses are in GeV.}
\label{Stopmodelparametersolution}
\begin{center}
\begin{tabular}{ c c c c } \hline \hline
  Parameter   & Value                     &  $50$ fb$^{-1}$ Stat.          & $100$ fb$^{-1}$ Stat.                  \\ \hline \hline
  $\alpha$    & $4.58$                    & $\pm 0.21$                     & $\pm 0.14$                             \\  
  $m_{3/2}$   & $13717$                   & $\pm 688$                    & $\pm 517$                           \\ 
  $n_m$       & $0.106$                    &  $\pm 0.015$               & $\pm 0.015$                    \\  
  $n_H$       & $0.578$                    &  $\pm 0.095$                    & $\pm 0.091$                              \\  
  ${\rm tan}\beta$ & $28.76$              &  $\pm 1.65$                 & $\pm 1.36$                     \\  \hline \hline
\end{tabular}
\end{center}
\end{table}

Using the model parameters from Table \ref{Stopmodelparametersolution}, we calculate the relic density and we find:
\be
\Omega h^2 \,\, = \,\, 0.096 \, \pm \, 0.029 \,\,.
\ee
The relic density is determined to an accuracy of $31\%$.

\subsection{Determination of $\tilde{b}$ and $\tilde{t}$ Masses}

The observables $M^{\rm end}_{bW}$ and $M^{\rm end}_{jW}$ are used to determine the third generation squark masses, once the other masses have been obtained with the observables previously determined.

Theoretically, the functional dependences are $M_{bW} = M_{bW}(m_{\tilde{b}},m_{\tilde{t}},m_{\tilde{g}})$ and $M_{jW} = M_{jW}(m_{\tilde{b}}, m_{\tilde{t}}, m_{\neu{1}})$. As before, the masses are varied independently around the benchmark point, and the collider experiment and determination of $M^{\rm end}_{bW}$ and $M^{\rm end}_{jW}$ is performed each time.

We show the masses of the third generation squarks with uncertainties in Table \ref{Stopthirdgenmasses}. The statistical uncertainties range between $2.5\%-11.3\%$.

\begin{table}[!htp] 
\caption{Solution to stop and sbottom masses at $50$ fb$^{-1}$ for a stop coannihilation benchmark point. All masses are in GeV.}
\label{Stopthirdgenmasses}
\begin{center}
\begin{tabular}{c c c c } \hline \hline
  Particle      & Mass  & $50$ fb$^{-1}$ Stat. & $100$ fb$^{-1}$ Stat.     \\ \hline \hline
  $\tilde{b}$   & 531   & $-$60, +60             & $-$47, +47                \\  
  $\tilde{t}$   & 326   & $-$5,  +8              & $-$4,  +7                \\ \hline \hline
\end{tabular}
\end{center}
\end{table}

\subsection{Gaugino Unification}\label{gauginounifforstop}

Upto this point, we have displayed our techniques of reconstructing masses at the benchmark point given in Table \ref{stopspectrumA}. Our techniques work perfectly well at benchmark points with higher gluino mass, as preferred by current LHC data. Higher luminosity is of course required to obtain endpoints. Below, we choose such a benchmark point with $m_{\sg} \sim 1.2$ TeV, reconstruct gaugino masses, and show the gaugino unification scale.

\begin{table}[!htp] 
\caption{Model parameters and spectrum at a new stop coannihilation benchmark point with heavier gluino. All masses are in GeV.}
\label{stopparametersB}
\begin{center}
\begin{tabular}{c c c c} \hline \hline
  Parameter      & Value    & Particle      & Mass     \\ \hline \hline
  $\alpha$       & 3.8      & \sg           & 1187     \\  
  $m_{3/2}$      & 34800    & \neu{2}       & 740            \\ 
  $n_m$          & 0.0      & \neu{1}       & 666             \\  
  $n_H$          & 0.5      & \stau         & 721           \\  
  ${\rm tan}\beta$ & 28     & \sq           & 1189      \\  \hline \hline
\end{tabular}
\end{center}
\end{table}

For this benchmark point, a luminosity of $200$ fb$^{-1}$ is required to solve for all the masses, following the techniques we have shown in this paper. We show the masses we obtained for this benchmark point in Table \ref{stopmasseshigherpoint}. The statistical uncertainties range between $0.9\%-14.7\%$.

\begin{table}[!htp] 
\caption{Solution to masses at $200$ fb$^{-1}$ for a stop coannihilation benchmark point with heavier gluino. All masses are in GeV.}
\label{stopmasseshigherpoint}
\begin{center}
\begin{tabular}{c c c c} \hline \hline
  Particle      & Mass      & Stat.          \\ \hline \hline
  $\st$         & 690       & $\pm$ 6             \\  
  $\tilde{b}$   & 1002      & $\pm$ 126          \\ 
  $\stau$       & 717       & $\pm$ 10           \\  
  $\sq$         & 1133      & $-132,+167$         \\  \hline \hline
\end{tabular}
\end{center}
\end{table}

The model parameters are solved as before, using the masses above as well as the gaugino masses solved at $200$ fb$^{-1}$. The solution to the model parameters at $200$ fb$^{-1}$ for the new benchmark point are $\alpha = 3.84 \pm 0.11, m_{3/2} = 34463 \pm 805 \,\,{\rm GeV}, n_m = 0.078 \pm 0.007, n_H = 0.592 \pm 0.036,$ and ${\rm tan} \beta = 27.2 \pm 1.94$. The relic density is found to be $\Omega h^2 = 0.23 \pm 0.13$.

As mentioned in Section \ref{intro}, gaugino masses may be obtained at lower luminosity, as we elaborate below.

The solution of gaugino masses typically requires the values of four observables ($p_{T,{\rm diff}}, p_{T,{\rm AM}}, M^{\rm peak}_{\rm eff},$ and $M^{\rm end}_{\tau \tau}$), as is clear from the solutions we displayed in Table \ref{Stopmasssolutions}. 
%
%
One could have used the endpoint of the invariant mass distribution $M_{j\tau}$ of the leading jet and the more energetic $\tau$ as an observable. Our analysis of this observable indicates that one requires more luminosity to find its endpoint, relative to the luminosity required to determine the observables $p_{T,{\rm AM}}$ and $p_{T,{\rm diff}}$, because of the associated jet subtractions. 

Since for the benchmark point the $\tau$'s are quite close in mass, we use $p_{T,{\rm AM}}$ and $p_{T,{\rm diff}}$, obviating the need to use observables which involve jet information. Then, $\neu{1}, \neu{2},$ and $\tilde{\tau}$ masses can be obtained from $p_{T,{\rm AM}}, p_{T,{\rm diff}},$ and $M^{\rm end}_{\tau\tau}$ observables. Since these observables do not involve jet subtractions, they give acceptable endpoints at lower luminosity. On the other hand, $M^{\rm peak}_{\rm eff}$ roughly gives the gluino mass.
 
In Table \ref{Stopgauginomasses}, we show the values of the gaugino masses at $50$ fb$^{-1}$ for the new benchmark point. The statistical uncertainties range between $1.2\%-4.2\%$.

\begin{figure}[!htp]
\centering
\includegraphics[width=3.5in]{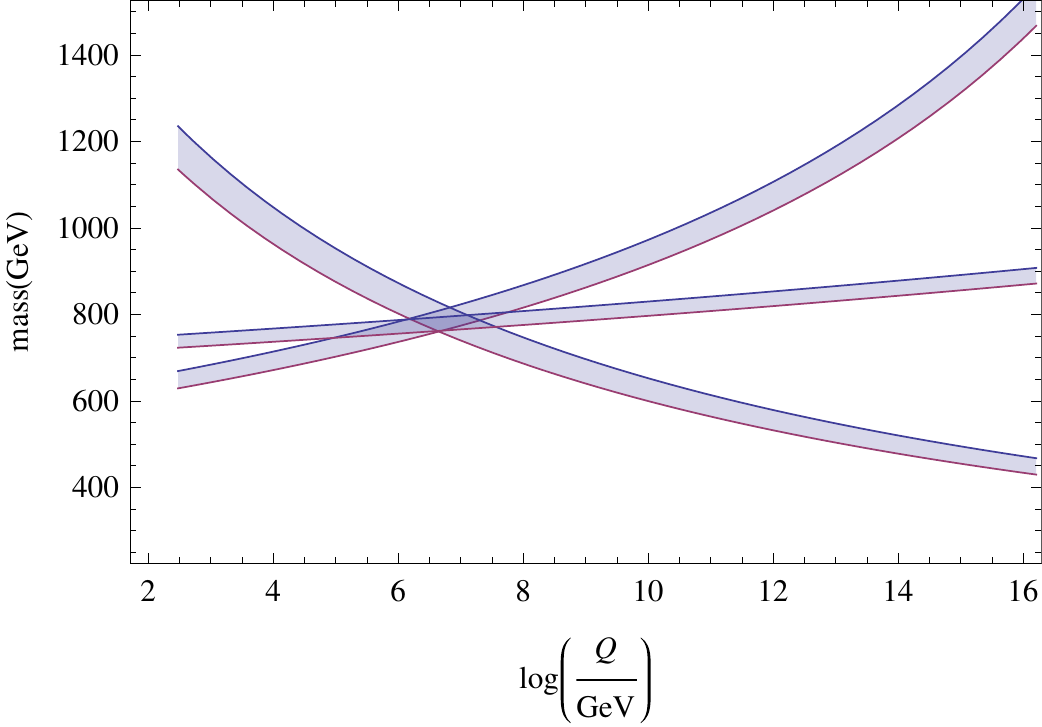}
\caption{Running of gaugino masses. The vertical axis is masses in GeV. The luminosity is $50$ fb$^{-1}$. The shaded lines from the top down represent, at the left edge of the graph, the gluino, wino and bino masses respectively.}
\label{Stopgauginounif}
\end{figure}

In Figure \ref{Stopgauginounif}, we show the running of the gauginos and the gaugino unification for our benchmark point. Clearly, gaugino unification occurs below the GUT scale, establishing the footprints of anomaly contributions to supersymmetry breaking.

\begin{table}[!htp] 
\caption{Solution to gaugino masses at $50$ fb$^{-1}$ for the new stop coannihilation benchmark point with heavier gluino. All masses are in GeV.}
\label{Stopgauginomasses}
\begin{center}
\begin{tabular}{c c c } \hline \hline
  Particle      & Mass        & $50$fb$^{-1}$ Stat.    \\ \hline \hline
  $\sg$         & 1181        & $\pm$ 50                          \\  
  $\neu{2}$     & 738         & $\pm$ 15                         \\ 
  $\neu{1}$     & 649         & $\pm$ 20                      \\  \hline \hline
\end{tabular}
\end{center}
\end{table}

%
%

The theoretically expected gaugino unification scale, for this benchmark point, is given by
\be
M_{\rm unif} \, = \, M_{GUT}\left(\frac{m_{3/2}}{M_p}\right)^{2.47/\alpha} \sim 10^7 \, {\rm GeV} \,\,\,.
\ee
Note that the expression differs from the ones in \cite{Choi:2007ka} by the rescaling in Eq.~\ref{rescaledalpha}.

\subsection{Systematic Errors}

We also study an impact due to a $\pm 3\%$ uncertainty on the energy scale for jets, taus, and missing transverse energy to estimate the systematic uncertainties independently of luminosity. 

Table \ref{Stopsystematicerrors} shows the percentage systematic errors in the determination of the observables and masses. We find that the systematic errors for the observables are between $1\%-6\%$, which is larger than the statistical error $0.2\%-3\%$ at $50$ fb$^{-1}$. The systematic errors of masses, which vary from $2\%-19\%$, are also large compared to the statistical errors, which range between $1\%-11\%$ at $50$ fb$^{-1}$. Since our assumption of $\pm 3\%$ on the energy scale could be changed in actual experimental conditions, we just show the size of errors in Table \ref{Stopsystematicerrors} as a reference and don't use them in the analyses performed in the previous sub-sections.

\begin{table}[!htp] 
\caption{Systematic percentage errors in observables and masses for a benchmark point in the stop coannihilation region.}
\label{Stopsystematicerrors}
\begin{center}
\begin{tabular}{c c c c c c} \hline \hline
  Observable                      & Error (\%)   & Mass              & Error (\%)      & Parameter           & Error (\%)\\ \hline \hline
  $M^{\rm end}_{\tau \tau }$      & $\pm$ 3.0      & $m_{\sg}$         & $\pm$ 3.9         & $\alpha$            &  $\pm$ 5.9 \\  
  $M^{\rm end}_{j\tau \tau }$     & $\pm$ 4.5      & $m_{\neu{2}}$     & $-$2.4,+4.2       & $m_{3/2}$           &  $\pm$ 5.9     \\ 
  $slope(p_{T,{\rm high}})$       & $\pm$ 5.2      & $m_{\neu{1}}$     & $\pm$ 4           & $n_m$               &  $\pm$ 12.5   \\  
  $slope(p_{T,{\rm low}})$        & $\pm$ 1.0      & $m_{\stau}$       & $\pm$ 2.6         & $n_H$               &  $\pm$ 17  \\
  $M^{\rm peak}_{\rm eff}$            & $\pm$ 1.3      & $m_{\sq}$         & $-$13,+15     & ${\rm tan} \beta$   &  $\pm$ 6.1    \\
  $M^{\rm end}_{jW}$              & $\pm$ 5.9      & $m_{\st}$         & $-$2.5,+3.7       &             &     \\
  $M^{\rm end}_{bW}$              & $\pm$ 3.7      & $m_{\tilde{b}}$   & $\pm$ 19          &             &     \\ \hline \hline
\end{tabular}
\end{center}
\end{table}


\section{Analysis in the Stau Coannihilation Region}\label{staucoannihilation}

In this Section, we determine masses and model parameters for a different point in parameter space, where only stau coannihilation is operational. We first show our analysis for the benchmark model point shown in Table \ref{Staubenchmarkpoint}. In Section \ref{gauginounifforstau}, we will also analyse a benchmark point with heavier gluino mass preferred by current LHC data.

The distributions used in the stau coannihilation region are $M_{\tau\tau}$, $M_{j\tau\tau}$, $M_{j\tau}$, $p_{T,{\rm low}}$, and $M_{\rm eff}$. Note that we only use the $p_{T,{\rm low}}$ corresponding to the lower energy $\tau$ from $\tilde{\tau} \longrightarrow \neu{1} + \tau$, since the higher energy $\tau$ doesn't show enough variation as the masses are varied. 

We use $M_{j\tau}$, which is formed by combining the higher energy $\tau$ with a leading jet from the same event. OS$-$LS and BEST are performed as usual. Theoretically, one expects  $M_{j\tau }^{\rm end} \, = \, M_{j\tau }^{\rm end}(m_{\tilde{q_L}},m_{\neu{2}}, m_{\tilde{\tau}})$.

\begin{figure}[ht] 
\centering
\includegraphics[width=3.5in]{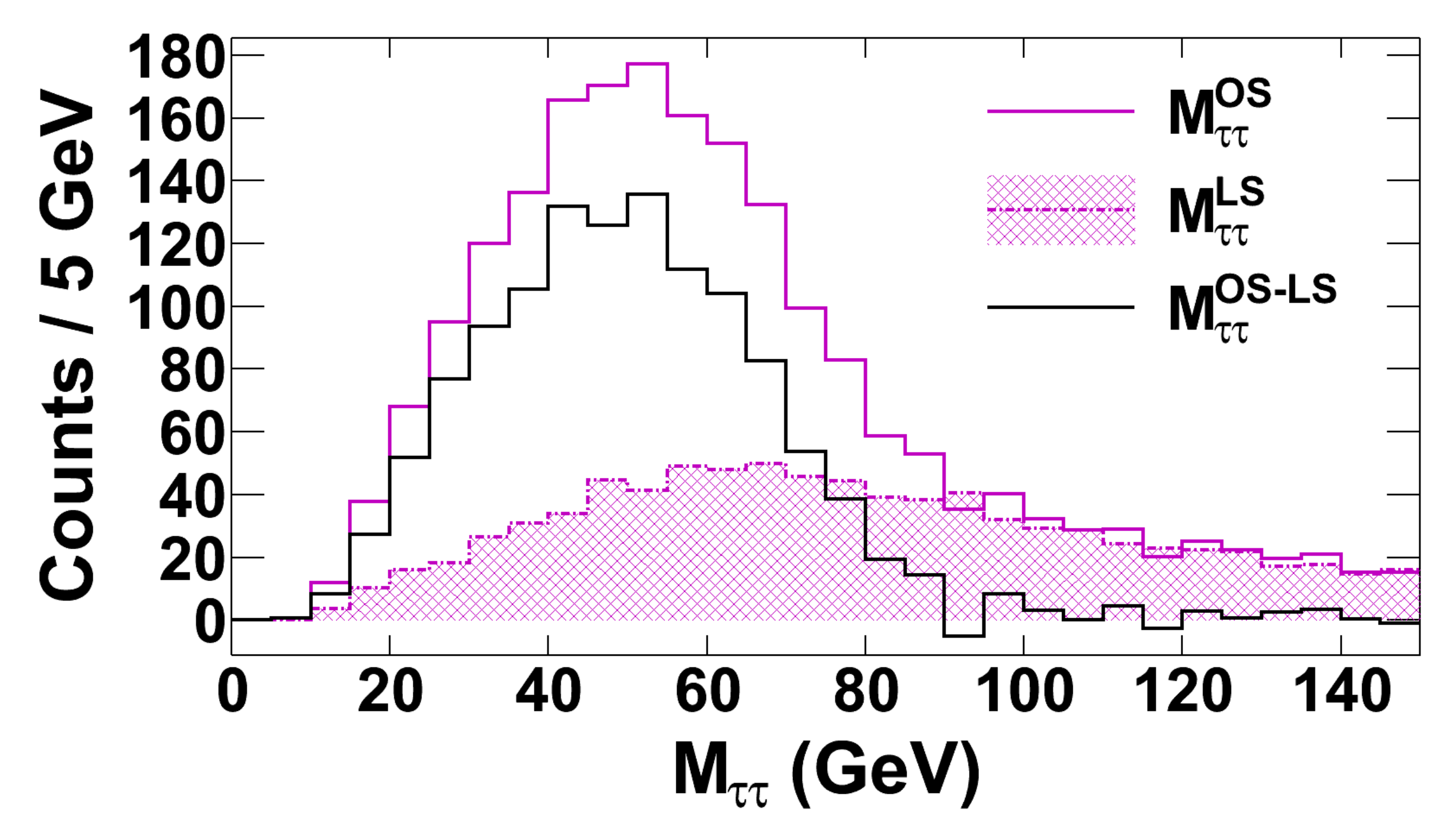}
\caption{Distribution of $M_{\tau \tau }$ at a stau coannihilation benchmark point. The notation is the same as in Figure \ref{StopMtt}. The endpoint obtained is $90.70 \pm 0.54 (\rm Stat.)$ GeV. The luminosity is $100$ fb$^{-1}$.}
\label{StauMtautau}
\end{figure}

\begin{figure}[ht] 
\centering
\includegraphics[width=3.5in]{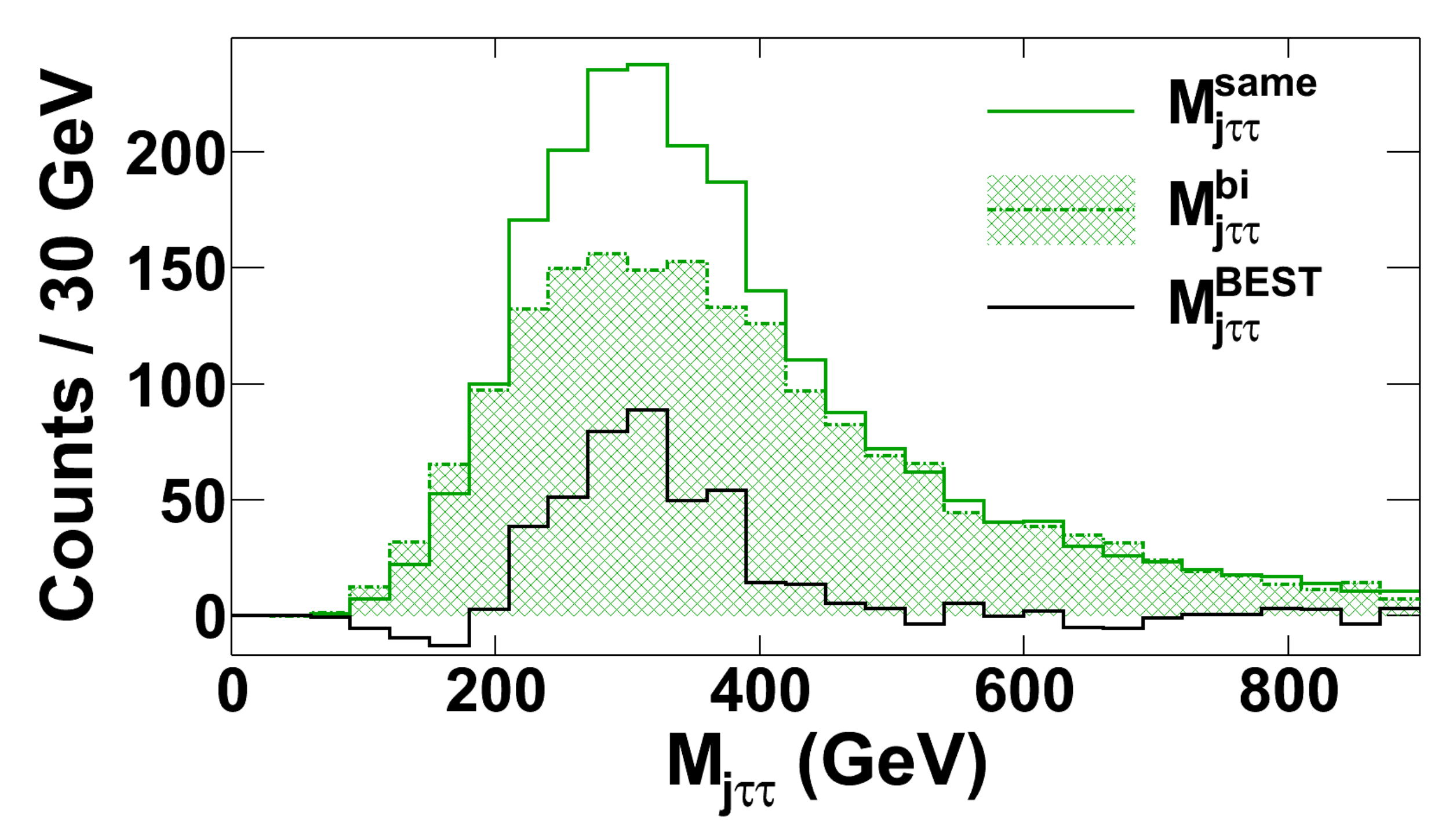}
\caption{Distribution of $M_{j\tau \tau }$ at a stau coannihilation benchmark point. The notation is the same as in Figure \ref{StopMjtt}. The endpoint obtained is $479.53 \pm 3.45 (\rm Stat.)$ GeV.  The luminosity is $100$ fb$^{-1}$.}
\label{Staujettautau}
\end{figure}

\begin{figure}[ht]
\centering
\includegraphics[width=3.5in]{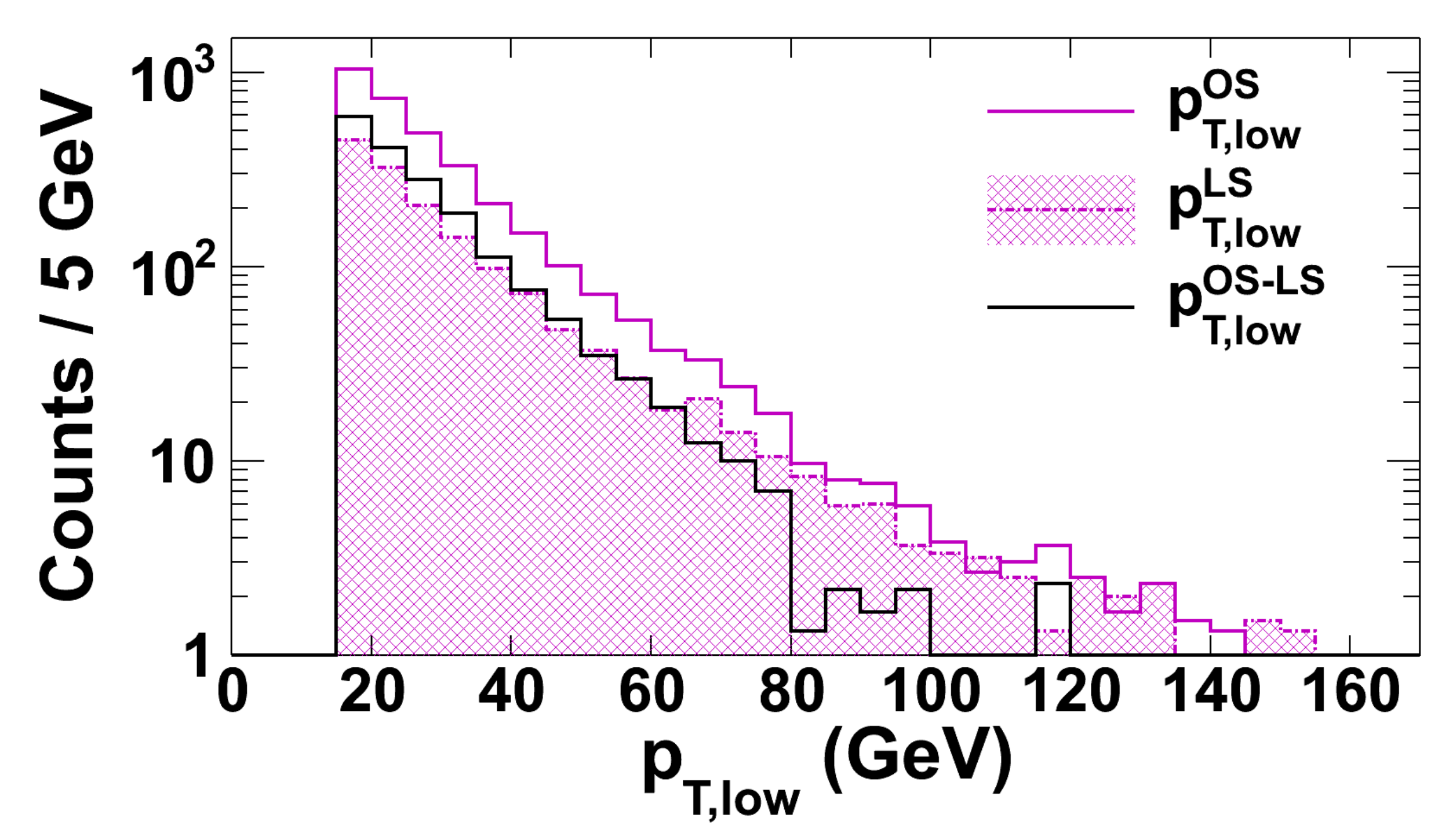}
\caption{Distribution of  $p_{T,{\rm low}}$ at a stau coannihilation benchmark point. The notation is the same as in Figure \ref{Stopptmin}. The slope obtained is $-0.0849 \pm 0.0041 (\rm Stat.)$. The luminosity is $100$ fb$^{-1}$.}
\label{StauPTtau}
\end{figure}

\begin{figure}[ht]
\centering
\includegraphics[width=3.5in]{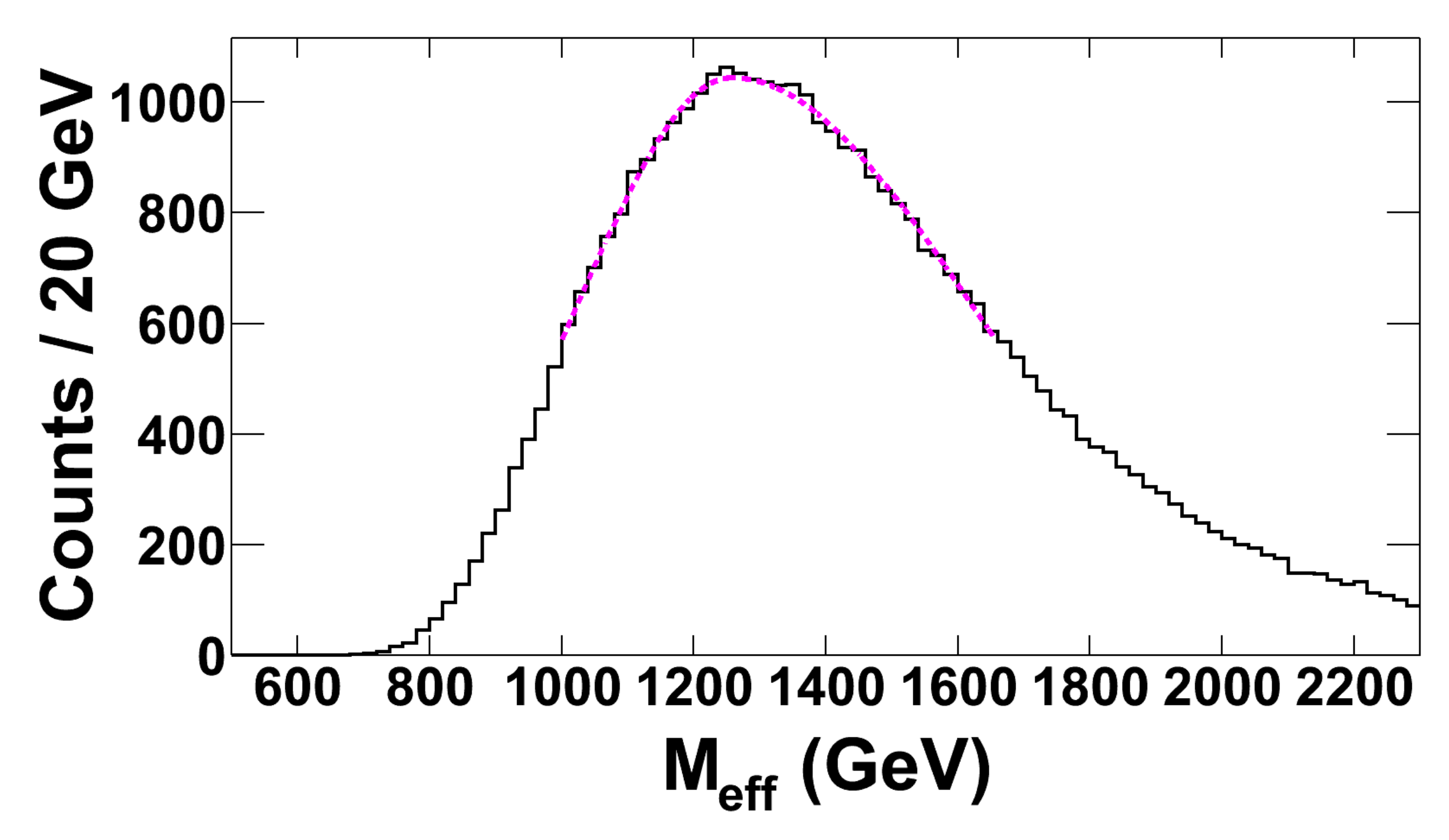}
\caption{Distribution of $M_{\rm eff}$ at a stau coannihilation benchmark point. The notation is the same as in Figure \ref{Stopmeff}. The peak obtained is $1257.26 \pm 10.33 (\rm Stat.)$ GeV. The luminosity is $100$ fb$^{-1}$.}
\label{StauMeff}
\end{figure}

\begin{figure}[ht]
\centering
\includegraphics[width=3.5in]{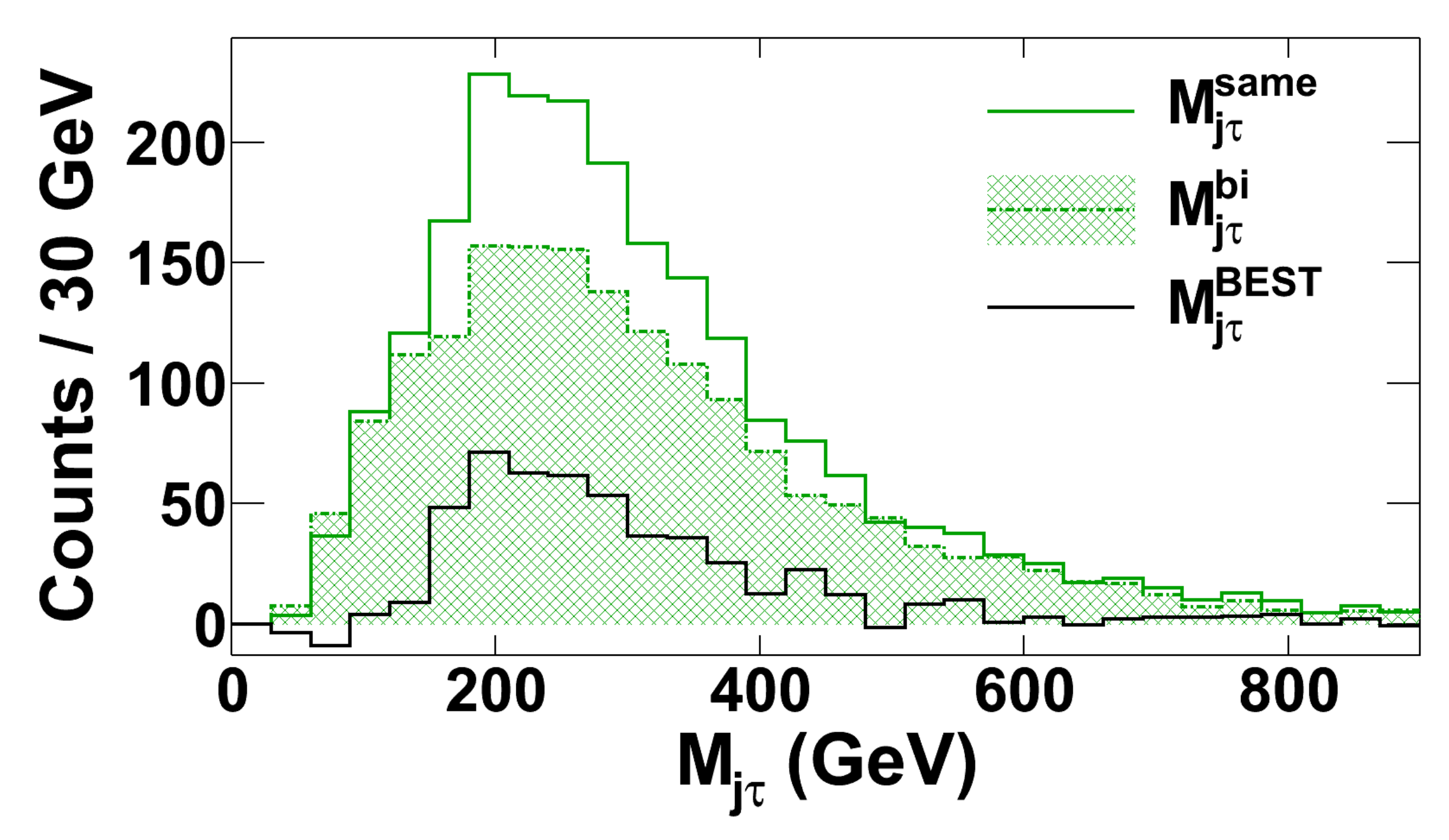}
\caption{Distribution of $M_{j\tau}$ in $j\tau\tau$ system. The blue (grey) histogram is obtained by combining the higher energy $\tau$ of the OS$-$LS $\tau$ pair with a leading jet of the same event. The filled dot-dashed histogram is the bi-event histogram, constructed by combining the aforesaid $\tau$ with a leading jet from a different event and is normalised to the long tail of the same-event histogram. The same-event minus bi-event subtraction (BEST) produces the black subtracted histogram. The subtracted histogram is fitted with a straight line to obtain the endpoint. The result for the endpoint is  $448.40 \pm 16.20 (\rm Stat.)$ GeV. The luminosity is $100$ fb$^{-1}$.}
\label{Staujettau}
\end{figure}

In Figures \ref{StauMtautau}, \ref{Staujettautau}, \ref{StauPTtau}, \ref{StauMeff}, and \ref{Staujettau}, we present the distributions obtained.

In Table \ref{Stauobservables100}, we present the endpoints obtained from the $M_{\tau\tau}, M_{j\tau\tau},$ and $M_{j\tau}$ distributions, the slope of the visible $p_T$ distribution of the lower energy $\tau$, and the peak of the $M_{\rm eff}$ distribution, at $100$ fb$^{-1}$ for the stau coannihilation benchmark point. The statistical uncertainties range between $0.6\%-4.7\%$.

\begin{table}[!htp]
\caption{Kinematical observables at $100$ fb$^{-1}$ for the Stau coannihilation benchmark point. All masses in GeV.}
\label{Stauobservables100}
\begin{center}
\begin{tabular}{c c c } \hline \hline
  Observable                        & Value       & $100$ fb$^{-1}$ Stat.               \\ \hline \hline
  $M^{\rm end}_{\tau \tau }$                  & $90.70$     & $\pm 0.54$                           \\  
  $M^{\rm end}_{j\tau \tau }$                 & $479.53$    & $\pm 3.45$                           \\ 
  $slope(p_{T,\tau})$                & $-0.0849$   & $\pm 0.0041$                         \\  
  $M^{\rm peak}_{\rm eff}$                         & $1257.26$   & $\pm 10.33$                        \\
  $M^{\rm end}_{j\tau}$                       & $448.40$    & $\pm 16.20$                        \\\hline \hline  
\end{tabular}
\end{center}
\end{table}


\subsection{Masses and Gaugino Unification}\label{staumassgaugunifsection}

After obtaining the values of the observables at the benchmark point as quoted above, we varied the masses around the benchmark point. For each mass selection, we simulated the experiment and obtained all the observables. This gave us the observables as a function of the masses. We then iteratively solved for the masses.

Obviously, one should expect more than one set of mass solutions in this method, since the dependence of the observables on the masses near the benchmark point is non-linear. In this case, we found two sets of solutions with quite disparate values of masses. 

This is a general problem, and we mention some arguments to reject the "wrong" solutions. Firstly, one set of solutions had masses far away from the benchmark point, where the expansion was not performed to begin with. Secondly, this set had a much larger stau-neutralino mass difference, which did not satisfy the relic density constraint. 

Finally, one can reject masses by appealing to the UV model and determining more observables, although this goes somewhat against the bottom-up philosophy we have pursued in this paper. Thus, we can use this set of mass solutions and solve for the UV model parameters, which we can then use to solve the spectrum of the model, in particular obtaining the $b$ quark mass. Given the spectrum, we can obtain the theoretical value of a particular observable, say $M_{bW}$, and on the other hand, measure it using the techniques outlined in the paper. For the incorrect mass solution set, the values thus obtained will be very different.

In any case, after rejecting the first set of solutions, the second set itself had two sets of values of $\neu{1}$ and $\tilde{\tau}$ masses, which were close to each other. One set of solutions is shown in Table \ref{Staumasssolutions}. The other solutions to the $\neu{1}$ and $\tilde{\tau}$ masses are $m_{\neu{1}} = 294$ GeV and $m_{\tilde{\tau}} = 337$ GeV. All combinations of masses (four in all) are used to solve the model parameters, as we outline in the next subsection.

We display the solution set in Table \ref{Staumasssolutions}, obtained at luminosity $100$ fb$^{-1}$. The statistical uncertainties are $2.7\%-6.4\%$.

\begin{table}[!htp] 
\caption{Solution to masses at $100$ fb$^{-1}$ for the stau coannihilation benchmark point. All masses are in GeV.}
\label{Staumasssolutions}
\begin{center}
\begin{tabular}{c c c } \hline \hline
  Particle      & Mass        & $100$ fb$^{-1}$ Stat.              \\ \hline \hline
  $\sg$         & 895         & $-35,+50$              \\  
  $\tilde{q}_L$ & 845         & $-36,+24$            \\ 
  $\neu{2}$     & 388         & $-9,+25$            \\  
  $\tilde{\tau}$ & 298        & $-8,+8$      \\  
  $\neu{1}$     & 274         & $-10,+10$       \\\hline  \hline
\end{tabular}
\end{center}
\end{table}

%









\subsection{Determination of UV Model Parameters and Relic Density}\label{modelparameters}


Having determined the masses of $\tilde{g}, \neu{1}, \neu{2}, \tilde{\tau}$, and $\tilde{q}$, we can use these to numerically solve for the parameters of the full model.

We use the Nelder-Mead method to do so. An initial simplex is chosen on the parameter space $\{m_{3/2}, \, \alpha, \, n_m, \, n_H, \, {\text {tan}}\beta,\}$ and \isajet \ is used to find the corresponding mass. The scan over parameter space proceeds until the masses calculated by \isajet \ are close to the solved values.

In the previous subsection, we have discussed two solutions each for $\neu{1}$ and $\tilde{\tau}$. We solved the model parameters for each combination of masses (four in all), and took the average of the values of the model parameters thus obtained. These average values are displayed in Table \ref{Staumodelparametersolutions}. The statistical uncertainties are $9\%-40\%$.

\begin{table}[!htp] 
\caption{Solution to model parameters at the stau coannihilation benchmark point. For each mass solution, the model parameters are solved, and the average of all the different sets of solutions is shown in the table. Masses are in GeV.}
\label{Staumodelparametersolutions}
\begin{center}
\begin{tabular}{c c c} \hline \hline
  Parameter      & Value       &   Stat.           \\ \hline \hline
  $\alpha$       & 7.42         & $\pm$ 0.58         \\  
  $m_{3/2}$      & 10171       & $\pm$ 882         \\ 
  $n_m$          & 0.52        & $\pm$ 0.09         \\  
  $n_H$          & 1.17         & $-0.07,+0.22$        \\  
  ${\rm tan}\beta$ & 33.1      & $\pm$ 7.8         \\  \hline \hline
\end{tabular}
\end{center}
\end{table}

Using the model parameters from Table \ref{Staumodelparametersolutions}, we calculate the relic density and we find:
\be
\Omega h^2 \,\, = \,\, 0.17_{-0.13}^{+0.12} \,\,.
\ee
The relic density is thus determined to an accuracy of $59\%$.


\subsection{Gaugino Unification}\label{gauginounifforstau}

Although upto this point, we have displayed our results in the stau coannihilation region at the benchmark point given in Table \ref{Stauspectrum}, our methods are applicable at other benchmark points with higher gluino mass, preferred by current LHC data. We display such a benchmark point for stau coannihilation below, and proceed to show gaugino unification.

\begin{table}[!htp] 
\caption{Model parameters and spectrum at a new stau coannihilation benchmark point with heavier gluino. All masses are in GeV.}
\label{stopparametersB}
\begin{center}
\begin{tabular}{c c c c} \hline \hline
  Parameter      & Value    & Particle      & Mass     \\ \hline \hline
  $\alpha$       & 10      & \sg            & 1183     \\  
  $m_{3/2}$      & 9500    & \neu{2}        & 499            \\ 
  $n_m$          & 0.5      & \neu{1}       & 337             \\  
  $n_H$          & 1.0      & \stau         & 361           \\  
  ${\rm tan}\beta$ & 37     & \sq           & 1119      \\  \hline \hline
\end{tabular}
\end{center}
\end{table}

At this benchmark point, we found acceptable endpoints at $250$ fb$^{-1}$, and solved for the observables and masses. The masses are $m_{\stau} = 346 \, \pm \, 12$ GeV and $m_{\sq} = 1138 \, \pm \, 72$ GeV. The masses of the gauginos can also be determined, although we show their values at a lower luminosity below. After determining the masses, the observables can be solved and they are $\alpha = 9.95 \pm 0.99, \, m_{3/2} = 9636 \pm 910 \,\,{\rm GeV}, \, n_m = 0.56 \pm 0.09, \,n_H = 1.14_{-0.08}^{+0.20}, \, $ and ${\rm tan} \beta = 40.0 \pm 7.0$. The relic density for the heavy gluino point is found to be $\Omega h^2 = 0.05_{-0.04}^{+0.21}$.

The gaugino masses can be obtained at lower luminosity, as spelled out in Section \ref{intro}.

In the stau coannihilation region, one can use the proximity of the stau and lightest neutralino masses to solve gaugino masses using fewer observables, in particular avoiding the observable $M^{\rm end}_{j\tau}$, which typically requires higher luminosity due to jet subtractions. This is reminiscent of what we did in the previous sections with stop coannihilation, although there we used the proximity of $\tau$ $p_T$'s. 

We take approximately
\be
m_{\tilde{\tau}} \, \sim \,  m_{\neu{1}} + 15 \,\,\,.
\ee
Thus, assuming that $M^{\rm peak}_{\rm eff}$ is approximately determined by $ m_{\tilde{g}}$, it is clear from Eq. \ref{theoryfunctions} that one can solve for $ m_{\tilde{g}}, m_{\neu{2}}, m_{\neu{1}}$ from $M^{\rm peak}_{\rm eff}, p_{T,{\rm low}}$ and $M_{\tau \tau }$. 

These observables have acceptable endpoints and peaks down to $15$ fb$^{-1}$. The main difference from the full set of observables is the $M_{j\tau}$ distribution, which requires a higher luminosity to obtain acceptable endpoints due to jet subtractions. 



In Table \ref{Staugauginomasses15}, we show the gaugino masses obtained at $15$ fb$^{-1}$ for the new stau coannihilation benchmark point with heavier gluino. The statistical uncertainties are $7\%-11.4\%$.

\begin{table}[!htp] 
\caption{Solution to gaugino masses at $15$ fb$^{-1}$ for the new stau coannihilation benchmark point with heavier gluino. All masses are in GeV.}
\label{Staugauginomasses15}
\begin{center}
\begin{tabular}{c c c } \hline \hline
  Particle      & Mass        & $15$fb$^{-1}$ Stat.   \\ \hline \hline
  $\sg$         & 1186         & $\pm$ 84         \\  
  $\neu{2}$     & 527         & $-30,+60$        \\ 
  $\neu{1}$     & 317         & $\pm$ 33        \\  \hline \hline
\end{tabular}
\end{center}
\end{table}

\begin{figure}[!htp]
\centering
\includegraphics[width=3.5in]{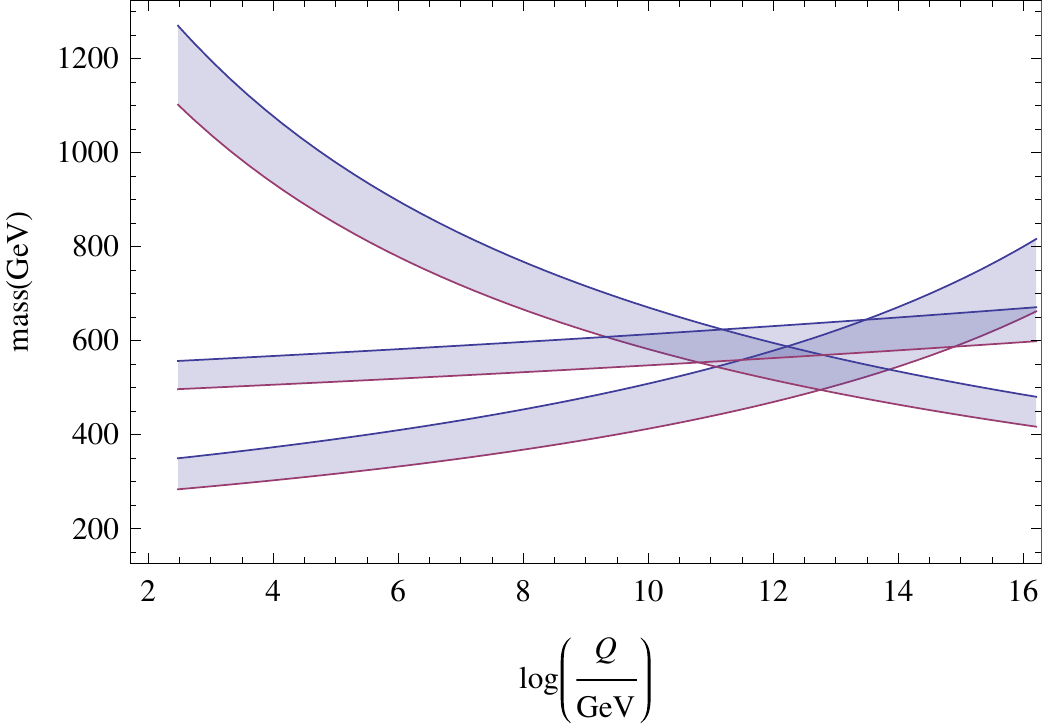}
\caption{Running of gaugino masses. The vertical axis is masses in GeV. The luminosity is $15$ fb$^{-1}$. The shaded lines from the top down represent, at the left edge of the graph, the gluino, wino and bino masses respectively.}
\label{agauginomasses15stau}
\end{figure}

In Figure \ref{agauginomasses15stau}, we show gaugino unification for the new benchmark point at $15$ fb$^{-1}$.

%
%



\section{Conclusions}

If supersymmetry is discovered at the LHC, the next step would be to understand as much as possible about the underlying model of supersymmetry breaking and mediation. How should one proceed?


One option is to start from specific models defined by a tractable number of parameters at high energy, and fit to observables measured in the collider. However, we have stressed that a given set of measured observables will generally point back at different UV models, distinguishing among which may be challenging and even intractable (depending on the complexity of the model and the number of reliable observables measured).

We have stressed on a bottom-up approach to this question: first identify sectors in the low energy spectrum which carry the imprints of mediation schemes in a relatively model-independent manner, next construct observables which carry information about these masses, and finally using the observables solve for the masses and identify a mediation scheme. 

The gaugino sector offers a particularly fruitful sector in this program, since the determination of this sector can unambiguously show the imprints of anomaly contributions to supersymmetry breaking. We therefore take an example of a model where anomaly contributions are significant (supersymmetry breaking in KKLT compactification) and proceed to obtain the gaugino masses, by constructing suitable observables. 

We rely on the construction of kinematic observables arising from cascade decays using endpoint techniques, since these are a particularly sharp tool for such diagnosis.

While the gauginos are important in the paradigm outlined above, they form only a subset of the masses we determined in this paper. Given a neutralino dark matter, the determination of other masses becomes a necessity when one wants to establish mechanisms to satisfy the relic density. This is perfectly amenable to the kind of bottom-up approach we have taken: in our examples, we needed to establish that the stop-neutralino and stau-neutralino coannihilation mechanisms were operational. To do so demanded the determination of stop and stau masses, in addition to the neutralino.

We summarize our work below:

$(i)$ We constructed observables using different combinations of the jets, $\tau$'s and $W$'s in the final states. We constructed these observables in the stop-neutralino and stau-neutralino coannihilation regions. 

In the stop coannihilation region, we constructed two new observables, $M^{\rm end}_{jW}$ and $M^{\rm end}_{bW}$, to determine the masses of the $\st$ and the $\tilde{b}$. The determination of the $\st$ and $\tilde{b}$ masses is especially challenging, since both decay to $b$ quarks. To make the problem worse, stop decays to missing energy by emitting a low energy jet in this coannihilation region. We showed how to reconstruct the stop mass in order to establish the fact that we are in the stop-neutralino coannihilation region.

We also constructed two other new observables, $p_{T,{\rm AM}}$ and $p_{T,{\rm diff}}$ which are combinations of the $p_T$ of two opposite sign $\tau$'s. These observables are particularly useful when the sample shows the presence of two opposite sign $\tau$'s that are close in mass.

$(ii)$ Using the observables, we solved for the masses of the gluino, the two lighter neutralinos, squarks (of the first two generations) and the lightest stau. We also determined the lighter stop and sbottom masses in the stop coannihilation region. 

$(iii)$ We used the neutralino and gluino masses to determine the gaugino unification scale and discern the effects of anomaly mediation from a bottom-up approach, as elaborated above.

$(iv)$ Using all the masses, we determined (a) the model parameters and (b) tested whether we are in a coanihilation region and dark matter relic density is satisfied.  We find that relic density can be determined with an accuracy of $31\%$ in the stop and $59\%$ in the stau coannihilation region.

As far as the diagnosis of the supersymmetry breaking mediation scheme is concerned, the most important future direction is obviously to expand our techniques to parts of the parameter space away from the coannihilation regions. Our observables enabled us to solve for the lightest neutralinos, and for the statements on mediation scheme that we made above, they were assumed to be gauginos (this is guaranteed in the coannhilation regions). It would be very interesting to test the paradigm outlined in this paper at regions where the Higgsino component is not insignificant. Also interesting would be to find bottom-up techniques to further distinguish between different schemes that correspond to the same gaugino mass pattern. We leave these questions for future work.




\section{Acknowledgements}

We would like to thank John Conley and Jamie Tattersall for discussions and Nathan Krislock for introducing us to the Nelder Mead method. K.W. would like to thank  Lingzhi Wang for help with programming. T.K. would like to thank D.H. Kim and Y. Oh for their constructive discussions throughout this work.
This work is supported in part by the DOE grant DE-FG02-95ER40917 and by the World Class University (WCU) project through the National Research Foundation (NRF) of Korea funded by the Ministry of Education, Science \& Technology (grant No. R32-2008-000-20001-0).








\end{document}